\documentclass[notitlepage,nofootinbib, reprint]{revtex4-1}
\pdfoutput=1
\usepackage[normalem]{ulem}
\usepackage{hyperref}
\usepackage{color}
\usepackage{amsmath, latexsym, amssymb}
\usepackage{amsfonts}
\usepackage{booktabs}
\usepackage{mathtools}
\usepackage{mathrsfs}

\definecolor{dodgerblue}{rgb}{0.12, 0.56, 1.0}
\definecolor{forestgreen}{rgb}{0.2, 0.6, 0.2}

\newcommand{\ud}{\mathrm{d}}
\newcommand{\uD}{\mathrm{D}}
\newcommand{\lie}{\pounds}
\newcommand{\bs}[1]{\boldsymbol{#1}}
\newcommand{\mc}[1]{\mathcal{#1}}
\newcommand{\ms}[1]{\mathscr{#1}}
\newcommand{\var}{\delta} 
\newcommand{\interchange}[2]{#1 \longleftrightarrow #2}
\newcommand{\nonoverlap}[2]{\mathchoice{#1(#2)}{#1(#2)}{#1\mathrlap{(#2)}}{#1\mathrlap{(#2)}}}
\newcommand{\V}[1]{\nonoverlap{V}{#1}}
\newcommand{\B}[1]{\nonoverlap{\mc B}{#1}}
\newcommand{\Sigmacap}[2]{\nonoverlap{\Sigma_{#1}}{#2}}
\newcommand{\ord}[2]{\underset{^{(#1)}}{#2}{}}
\newcommand{\<}{\langle}
\renewcommand{\>}{\rangle}
\newcommand{\flow}{\Upsilon}
\newcommand{\scri}{\ms{I}}

\DeclareMathOperator{\sgn}{sgn}

\begin{document}

\title{Flux-balance laws in scalar self-force theory}

\author{Alexander M.\ Grant}
\email{a.m.grant@soton.ac.uk}
\affiliation{Department of Physics, University of Virginia, P.O.~Box 400714,
  Charlottesville, Virginia 22904-4714, USA}
\affiliation{School of Mathematical Sciences, University of Southampton,
  Southampton, SO17 1BJ, United Kingdom}

\author{Jordan Moxon}
\affiliation{Theoretical Astrophysics, Walter Burke Institute for Theoretical Physics, California Institute of Technology, Pasadena, CA 91125, USA.}

\begin{abstract}
  The motion of a radiating point particle can be represented by a series of geodesics whose ``constants'' of motion evolve slowly with time.
  The evolution of these constants of motion can be determined directly from the self-force equations of motion.
  In the presence of spacetime symmetries, the situation simplifies: there exist not only constants of motion conjugate to these symmetries, but also conserved currents whose fluxes can be used to determine their evolution.
  Such a relationship between point-particle motion and fluxes of conserved currents is a \emph{flux-balance law}.
  However, there exist constants of motion that are not related to spacetime symmetries, the most notable example of which is the \emph{Carter constant} in the Kerr spacetime.
  In this paper, we first present a new approach to flux-balance laws for spacetime symmetries, using the techniques of symplectic currents and symmetry operators, which can also generate more general conserved currents.
  We then derive flux-balance laws for all constants of motion in the Kerr spacetime, using the fact that the background, geodesic motion is integrable.
  For simplicity, we restrict derivations in this paper to the scalar self-force problem.
  While generalizing the discussion in this paper to the gravitational case will be straightforward, there will be additional complications in turning these results into a practical flux-balance law in this case.
\end{abstract}

\maketitle

\tableofcontents

\section{Introduction}

True test-bodies in general relativity follow geodesics determined by the metric in a given spacetime.
This analysis, however, is an idealization: physical objects source gravitational fields themselves, changing the metric of the spacetime through which they are traveling.
At zeroth order in the mass of the body in question, it follows a geodesic, but the higher-order corrections are important for long-lived systems: these corrections are collectively known as the \emph{self-force}.

While self-force effects are typically neglected, one situation in which they are particularly relevant is the case of extreme mass-ratio inspirals (EMRIs).
These are systems characterized by a stellar-mass compact object (of mass $m$) orbiting a supermassive black hole (of mass $M \gg m$).
These systems emit gravitational waves in frequency ranges inaccessible to ground-based gravitational wave detectors (for example, due to seismic noise and arm length limitations), but will be detectable by space-based interferometers such as LISA~\cite{Amaro-Seoane2017, Babak2017, Gair2017}.
These systems present an entirely different regime in which to study gravitational waves: since the inspiral will last for years, and capture $\sim M/m$ orbits, the details of the waveform will provide detailed information about the spacetime of the supermassive black hole~\cite{Ryan1995, Babak2017}.
Moreover, not only first-order self-force, but second-order self force effects will be relevant for data analysis of EMRI waveforms detected by LISA~\cite{Hinderer2008, Isoyama2012}.

The self-force formalism itself has a long history; for a review, see~\cite{Poisson2011} and references therein.
For most of its history, much of the focus has been on the first-order self-force, although second (and higher) order effects have now been placed on a firm footing as well (see~\cite{Pound2015b} and references therein).
In this paper, we will take the self-force formalism as given, and will focus exclusively on \emph{solving} the equations.

The solution to the self-force equations of motion comes in two parts: the motion of the body, and the perturbations to the fields which propagate on the background spacetime.
For practical applications, it is only the latter that we fundamentally wish to compute: we want to know the asymptotic radiation that is emitted by the system and reaches our detectors.
Given the motion of the body, computing this radiation at first order only requires the first-order field asymptotically.
At second order, while the first-order field is required everywhere, the second-order field is still only required to be known asymptotically.

In contrast, the computation of the motion itself is difficult: it involves fields at the location of the object, not just asymptotically.
While black hole perturbation theory most easily produces components of the Weyl tensor (see the recent review~\cite{Pound2021}, and references therein), the self-force evolution of the motion requires the metric perturbation, and so a \emph{metric reconstruction} procedure is needed~\cite{Chrzanowski1975, Cohen1979}, unless the metric perturbation has been computed directly.
While the methods for reconstructing a perturbed metric are relatively well-understood at first order~\cite{Pound2013, Merlin2016, van_De_Meent2017} (for certain gauge choices), second-order metric reconstruction procedures are much more complicated and will require significantly more work~\cite{Toomani2021}.
Moreover, metric reconstruction is far easier in asymptotic regions far away from the object.

The goal of flux-balance laws is to short-circuit this unfortunate complication: instead of determining the motion of the object using local fields, flux-balance laws determine \emph{some aspect} of the motion using only asymptotic fields.
The intuitive picture for these flux-balance laws is one that often is used in electromagnetism (see, for example, Sec.~16.2 of~\cite{Jackson1999}): determining the change in energy of a charged particle only requires knowledge of the total power radiated, and not the details of the local electromagnetic forces that act upon it.
It is important, however, to note that this only holds in a time-averaged sense: at any instant of time, the electromagnetic field itself contains energy, some of which will be radiated off to infinity and some of which will return to the particle at a later point.

This intuitive picture of flux-balance laws, while useful, does not ultimately provide a firm foundation on which to base a calculation.
One needs to \emph{prove} that a flux-balance law holds, by relating the evolution of some property of an object's motion to the flux of a conserved current at infinity.
Such flux-balance laws were first developed in the Kerr spacetime, relating the changes in energy $E$ and the $z$ component of the angular momentum $L_z$ to fluxes at infinity (and, importantly, the horizon of the black hole!) by Gal'tsov~\cite{Galtsov1982}.
These relationships only held for changes in these quantities in the limit where one considered an infinite amount of time, but by dividing by the time difference before taking the limit, this provides a relationship for \emph{average} rates of change of these quantities.
A rigorous analysis~\cite{Quinn1999} later showed that such an analysis held for any spacetime possessing Killing vectors.
Mino~\cite{Mino2003} showed that \emph{orbital} averages of the changes in $E$ and $L_z$ could also be written as expressions involving only asymptotic fields at the horizon and infinity.

The results of Mino~\cite{Mino2003} were surprising, since they also contained an expression for the orbit-averaged change in the third constant of motion in the Kerr spacetime, the Carter constant $K$~\cite{Carter1968a}.
While the conserved currents used by~\cite{Galtsov1982, Quinn1999} for arbitrary Killing vectors were constructed from the stress-energy tensor of the theory in question (or effective stress-energy tensor, in the case of gravity), it was not known if conserved currents associated with Killing tensors (from which the Carter constant can be constructed~\cite{Walker1970}) could be constructed in a similar way.
Moreover, it was later shown that there could be \emph{no} conserved current that could be constructed from the stress-energy tensor and this Killing tensor, under the assumption that this conserved current reduced to the Carter constant of a point particle when evaluated using the point-particle stress-energy tensor~\cite{Grant2015}.
While there existed conserved currents associated with the Carter constant for scalar fields~\cite{Carter1977}, and these results were extended to the case of linearized gravity in~\cite{Grant2020}, it was not clear if these conserved currents could be used to generate a flux-balance law.

In this paper, we attempt to derive flux-balance laws using a class of conserved currents that was used in~\cite{Grant2020}.
Instead of arising from the stress-energy tensor, these conserved currents are defined from the \emph{symplectic current}, a bilinear conserved current that is defined for perturbations to a field theory defined from a Lagrangian~\cite{Burnett1990, Lee1990}, and symmetry operators, which are operators which map the space of solutions to a theory into itself~\cite{Miller1977, Kalnins1992}.
We first use these techniques in order to derive flux-balance laws for conserved quantities that arise from spacetime symmetries, such as $E$ and $L_z$.
For the Carter constant, however, it seems that, despite the existence of conserved currents related to the Killing tensor, these conserved currents do not provide flux-balance laws that determine the evolution of the Carter constant.

The failure of these conserved currents to determine the evolution of the Carter constant seems, in part, to be caused by the fundamental difference between the Carter constant and $E$ and $L_z$: as it is constructed from a rank two Killing tensor, it is \emph{quadratic} in the momentum of the particle.
As such, we are motivated to consider a formulation in which all of the constants of motion for geodesic motion are on equal footing: action-angle variables for the Hamiltonian formulation.
The four conserved quantities, $E$, $L_z$, $K$, and $m^2$, can be written in terms of a set of four action variables, which we collectively denote by $J_\alpha$.
The evolution of these action variables can then be written in a unified manner in terms of the Hamiltonian.

Somewhat miraculously, changes in the action variables can be understood in terms of a flux-balance law.
While flux-balance-like expressions for the action variables have appeared previously in~\cite{Isoyama2018}, here we show that there exist flux-balance laws that can be derived directly in terms of a conserved current.
This conserved current is generated using a different sort of symmetry operator for the fields in question: these operators take advantage of the fact that the fields that occur in this problem are not arbitrary, but are dependent upon some given worldline.
This worldline, in turn, depends on an initial point at some given time, and therefore these fields can be differentiated with respect to this initial point.
This differential operator provides a map from the space of solutions (dependent on a worldline) into itself, and so is a symmetry operator.
The resulting symmetry operator can be used to generate a conserved current, the flux of which, as we will show, determines the evolution of the action variables $J_\alpha$.

Since these flux-balance laws require only fields off the worldline (though, due to a caveat which we will discuss further in Sec.~\ref{sec:averaged}, they are not asymptotically defined fields), this flux-balance computation forms the basis for a potentially useful means of computing the evolution of the Carter constant.
Moreover, while the computations in this paper are entirely at first order, the generalization to the second-order self-force is expected to be simpler than generalizing the results of Mino~\cite{Mino2003} directly.

The layout of the rest of the paper is as follows: in Sec.~\ref{sec:eom}, we discuss the equations of motion of a point particle under the scalar self force, both in terms of the usual formulation in terms of the force that acts on the particle and in terms of a Hamiltonian formulation.
In particular, we discuss the quasi-conserved quantities that arise, and discuss how the self-force is described by a perturbed Hamiltonian system.
In Sec.~\ref{sec:flux_balance}, we then turn to the derivation of flux-balance laws for the scalar self-force.
To do so, we first discuss conserved currents that arise in this theory, in particular the symplectic current, and then derive a series of ``integrated'' flux-balance laws related to changes in quantities considered in Sec.~\ref{sec:eom}, as well as ``averaged'' flux-balance laws which capture their average evolution.
We provide our conclusions, and a roadmap for the gravitational case, in Sec.~\ref{sec:conclusions}.
We also include an example calculation in Appendix~\ref{app:radiative} which is motivated by the discussion in Sec.~\ref{sec:averaged}.

We use the following notation and conventions in this paper: following Wald~\cite{Wald1984}, we use the ``mostly plus'' metric signature convention and lowercase Latin letters ($a$, $b$, etc.) for abstract indices for tensor fields defined on the spacetime manifold $\mc M$, while for coordinate indices we use lowercase Greek letters ($\alpha$, $\beta$, etc.).
We also use the conventions for differential forms from Appendix B of~\cite{Wald1984}.
For tensor fields on phase space $T^* \mc M$, we use uppercase Latin letters ($A$, $B$, etc.) for abstract indices, and Hebrew letters ($\aleph$, $\beth$, etc.) for coordinate indices.
Typically, quantities on phase space that are related to quantities on the spacetime manifold are the uppercase versions thereof: for example, a curve $\gamma$ on the spacetime manifold is given by the projection of a curve $\Gamma$ through phase space.
For arbitrary collections of abstract tensor indices, we use uppercase script Latin letters ($\ms A$, $\ms B$, etc.).
Our notation for bitensors matches that of~\cite{Poisson2011}, and we use the convention that indices at a point $x$ with some adornments have the same adornments: for example, $a'$, $b'$, etc. denote indices at $x'$.
As such, we drop the explicit dependence of bitensors on points at which they are evaluated, unless it is a scalar at that point.
We occasionally drop indices (such as in the case where we are considering differential forms): in such cases, we denote the tensors in bold, and directly apply any adornments (such as primes) to the tensor itself.
Finally, we denote the arguments of (multi-)linear functionals with curly brackets, to distinguish them from general, nonlinear functionals (which are typically denoted with square brackets).

\section{Equations of motion} \label{sec:eom}

The body whose motion we wish to determine follows a curve $\gamma(\varepsilon)$, parameterized by proper time $\tau$, with a parameter $\varepsilon$ that we use to track the scale of small perturbations.
The fundamental equation that we are concerned with is the following:\footnote{Apart from a minus sign, this equation agrees with Eq.~17.50 of~\cite{Poisson2011}.
  This minus sign is such that Eq.~\eqref{eqn:ret_eom} does \emph{not} have a minus sign (compare with the discussion of scalar self force in~\cite{Harte2014}).
  Choosing to have or not have this minus sign is equivalent to changing the sign of the scalar charge in all expressions.
  Note that there should also be a factor of $4\pi$ present in Eq.~\eqref{eqn:ret_eom}, comparing to Eq.~12.1 of~\cite{Poisson2011}.
  By dropping this factor, we are choosing to use a rationalized system of units (see the preface to~\cite{Heaviside1894}) for the scalar field, and so this factor of $4\pi$ appears in the Green's functions $G^{\rm R} (x, x')$ and $G^+ (x, x')$, or generally in solutions to the field equations, instead of the field equations themselves.}
\begin{equation} \label{eqn:scalar_sf}
  \dot{\gamma}^b (\varepsilon) \nabla_b p_a (\varepsilon) = -\varepsilon q \nabla_a \phi^{\rm R} + O(\varepsilon^2),
\end{equation}
where
\begin{equation}
  p_a (\varepsilon) \equiv m (\varepsilon) g_{ab} \dot{\gamma}^b (\varepsilon),
\end{equation}
and
\begin{equation} \label{eqn:normalization}
  \dot{\gamma}^a (\varepsilon) \dot{\gamma}^b (\varepsilon) g_{ab} = -1.
\end{equation}
We use the overdot (as in $\dot{\gamma}^a$) to represent differentiation with respect to proper time.
Below, for brevity, we denote by $\gamma$ the ``background'' curve $\gamma(\varepsilon)|_{\varepsilon = 0}$, and more generally places where an expected $\varepsilon$ argument is dropped indicates that the equation holds when $\varepsilon = 0$.

The only two properties that we will assume for the scalar field $\phi^{\rm R}$ which appears in Eq.~\eqref{eqn:scalar_sf} is that it is a solution to the sourceless, massless scalar field equation,
\begin{equation} \label{eqn:reg_eom}
  \Box \phi^{\rm R} = 0,
\end{equation}
and that it can be constructed by a procedure similar to integrating a Green's function over $\gamma$:
\begin{equation} \label{eqn:reg_integral}
  \phi^{\rm R} (x) = \int_{\V{\infty, -\infty}} \bs \epsilon' G^{\rm R} (x, x') \rho(x'),
\end{equation}
where $\V{\tau', \tau}$ denotes a spacetime volume which contains $\gamma(\tau'')$ for $\tau'' \in [\tau, \tau']$ and is such that $\gamma(\tau), \gamma(\tau') \in \partial \V{\tau', \tau}$.
The volume 4-form is represented by $\bs \epsilon$ and is adorned according to the integration point.
Moreover, the density $\rho$ is defined by
\begin{equation}
  \rho(x) \equiv q \int_{-\infty}^\infty \ud \tau\; \delta[x, \gamma(\tau)].
\end{equation}
Here, the delta function is defined to be the distribution that satisfies
\begin{equation}
  \int_V \bs \epsilon' f(x') \delta(x, x') = \begin{cases}
    f(x) & x \in V \\
    0 & x \not \in V
  \end{cases}.
\end{equation}
We also consider a different scalar field $\phi^+$ which is the retarded solution to the massless scalar field equation, with source $\rho$:
\begin{equation} \label{eqn:ret_eom}
  \Box \phi^+ = \rho,
\end{equation}
and
\begin{equation} \label{eqn:ret_integral}
  \phi^+ (x) = \int_{\V{\infty, -\infty}} \bs \epsilon' G^+ (x, x') \rho(x'),
\end{equation}
where $G^+ (x, x') = 0$ if $x'$ is not in the past of $x$.

\subsection{Conserved quantities}

In the presence of a Killing vector $\xi^a$, the background curve $\gamma$ possesses a conserved quantity given by
\begin{equation}
  E_\xi \equiv \xi^a p_a.
\end{equation}
In the case where $\xi^a = -(\partial_t)^a$, this conserved quantity is the energy, while in the case where $\xi^a = (\partial_\phi)^a$, it is the $z$-component of the angular momentum.
We now define a ``conserved'' quantity $E_\xi (\tau, \varepsilon)$ by
\begin{equation}
  E_\xi (\tau, \varepsilon) \equiv \xi^a p_a (\varepsilon).
\end{equation}
By Killing's equation $\nabla_{(a} \xi_{b)} = 0$, this quantity satisfies
\begin{equation}
  \frac{\ud_\varepsilon E_\xi (\varepsilon)}{\ud \tau} = -\varepsilon q \lie_\xi \phi^{\rm R} + O(\varepsilon^2),
\end{equation}
where we use the notation $\ud_\varepsilon$ to remind the reader that this derivative is along the curve $\gamma(\varepsilon)$, not $\gamma$.
As such, the change in the conserved quantity $E_\xi (\tau, \varepsilon)$ can be written as an integral over the worldline $\gamma(\tau, \varepsilon)$:
\begin{equation} \label{eqn:Delta_E_xi_exact}
  \begin{split}
    \Delta E_\xi (\tau', \tau; \varepsilon) &\equiv E_\xi (\tau', \varepsilon) - E_\xi (\tau, \varepsilon) \\
    &= -\varepsilon q \int_\tau^{\tau'} \ud \tau'' \lie_\xi \phi^{\rm R} [\gamma(\tau'', \varepsilon)] + O(\varepsilon^2).
  \end{split}
\end{equation}

We now vary with respect to $\varepsilon$, which we denote with a $\var$: for any quantity $Q(\varepsilon)$,
\begin{equation} \label{eqn:var_def}
  \var Q \equiv \left.\frac{\ud Q}{\ud \varepsilon}\right|_{\varepsilon = 0}.
\end{equation}
We find therefore that
\begin{equation}
  \begin{split}
    \var \Delta E_\xi (\tau', \tau) &= -q \int_\tau^{\tau'} \ud \tau'' \lie_\xi \phi^{\rm R} [\gamma(\tau'')] \\
    &= -q \int_\tau^{\tau'} \ud \tau'' \int_{\V{\infty, -\infty}} \bs \epsilon''' \delta[x''', \gamma(\tau'')] \lie_\xi \phi^{\rm R} (x''').
  \end{split}
\end{equation}
Using the fact that
\begin{equation} \label{eqn:delta_switch}
  \begin{split}
    \int_\tau^{\tau'} &\ud \tau'' \int_{\V{\infty, -\infty}} \bs \epsilon''' \delta[x''', \gamma(\tau'')] f(x''') \\
    &= \int_{\V{\tau', \tau}} \bs \epsilon'''\;\, \int_{-\infty}^\infty \ud \tau'' \delta[x''', \gamma(\tau'')] f(x'''),
  \end{split}
\end{equation}
we can therefore write
\begin{equation} \label{eqn:Delta_E_xi}
  \var \Delta E_\xi (\tau', \tau) = -\int_{\V{\tau', \tau}} \bs \epsilon\, \rho \lie_\xi \phi^{\rm R}.
\end{equation}

Note, however, that there are more types of conserved quantities that can be defined in arbitrary spacetimes; particularly relevant for the Kerr spacetime are those defined using a rank two Killing tensor $K^{ab}$:
\begin{equation}
  Q_K \equiv K^{ab} p_a p_b.
\end{equation}
As before, we define $Q_K (\tau, \varepsilon)$ using $p_a (\varepsilon)$, and it follows from the rank two Killing tensor equation $\nabla_{(a} K_{bc)} = 0$ that
\begin{equation}
  \frac{\ud_\varepsilon Q_K (\varepsilon)}{\ud \tau} = -2 \varepsilon q \lie_{K \cdot p(\varepsilon)} \phi^{\rm R} + O(\varepsilon^2),
\end{equation}
where
\begin{equation}
  (K \cdot p)^a (\varepsilon) \equiv K^{ab} p_b (\varepsilon).
\end{equation}
A similar set of steps as above shows that
\begin{equation} \label{eqn:Delta_Q_K}
  \var \Delta Q_K (\tau', \tau) = -2 \int_{\V{\tau', \tau}} \bs \epsilon\, \rho \lie_{K \cdot p} \phi^{\rm R},
\end{equation}
where $\Delta Q_K (\tau', \tau)$ is defined in a manner analogous to Eq.~\eqref{eqn:Delta_E_xi_exact}.
Note that this equation is well-defined, since the presence of $\rho$ in the integrand implies that the integrand has support only on the worldline (where $p_a$ is defined).

\subsection{Hamiltonian formulation}

Writing coordinates on the manifold $\mc M$ as $x^\alpha$, and considering momenta $p_\alpha$, we write coordinates $X^\aleph$ on phase space $T^* \mc M$ (the cotangent bundle) as
\begin{equation}
  X^\aleph \equiv \begin{pmatrix}
    x^\alpha \\
    p_\alpha
  \end{pmatrix}.
\end{equation}
As remarked in the introduction, we use Hebrew letters for coordinate indices on phase space.

For simplicity, we show that the self-force problem has a Hamiltonian form in these coordinates.
Consider the following Hamiltonian\footnote{Note that, in this Hamiltonian, we are considering the dependence on the worldline in $\phi^{\rm R}$ to be \emph{fixed}: while constructing Hamilton's equations, we do not vary this worldline $\gamma$, and then set the value of $x$ to be along the curve $\gamma$ at the very end of the calculation.
  As such, this is not truly a Hamiltonian system in the strict sense: see the discussion in~\cite{Vines2015, Fujita2016, Blanco2022} (as mentioned in~\cite{Blanco2022}, this discussion applies for scalar, electromagnetic, and gravitational self force).
  However, this will not affect the analysis of this paper, and so we drop the explicit dependence of $\phi^{\rm R}$ on $\gamma$ for brevity.}:
\begin{equation} \label{eqn:hamiltonian}
  H(X, \varepsilon) \equiv -\sqrt{-g^{\alpha\beta} (x) p_\alpha p_\beta} + \varepsilon q \phi^{\rm R} (x) + O(\varepsilon^2).
\end{equation}
Hamilton's equations then take the form
\begin{subequations} \label{eqn:hamilton}
  \begin{align}
    \frac{\ud x^\alpha}{\ud \tau} &= \frac{g^{\alpha\beta} p_\beta}{\sqrt{-g^{\gamma\delta} p_\gamma p_\delta}} + O(\varepsilon^2), \\
    \frac{\ud p_\alpha}{\ud \tau} &= -\frac{1}{2} \frac{p_\beta p_\gamma \partial_\alpha g^{\beta\gamma}}{\sqrt{-g^{\delta\epsilon} p_\delta p_\epsilon}} - \varepsilon q \partial_\alpha \phi^{\rm R} + O(\varepsilon^2).
  \end{align}
\end{subequations}
Note that the first of these equations implies Eq.~\eqref{eqn:normalization}, while the second, using
\begin{equation}
  \partial_\alpha g^{\beta\gamma} = -2 \Gamma^{(\beta}{}_{\delta\alpha} g^{\gamma)\delta},
\end{equation}
implies Eq.~\eqref{eqn:scalar_sf}.

Hamilton's equations can be recast in a covariant form on phase space by constructing a \emph{symplectic two-form} $\Omega_{AB}$, where we use capital Latin letters as abstract indices on phase space:
\begin{equation} \label{eqn:canonical}
  \bs \Omega \equiv \ud p_\alpha \wedge \ud x^\alpha.
\end{equation}
Note, in particular, that $\Omega_{AB}$ is a closed two-form:
\begin{equation} \label{eqn:d_Omega}
  \ud \bs \Omega = 0,
\end{equation}
because it is also an exact form, being the exterior derivative of the \emph{canonical one-form} $\bs \Theta$:
\begin{equation}
  \bs \Omega = \ud \bs \Theta,
\end{equation}
where
\begin{equation}
  \bs \Theta \equiv p_\alpha \ud x^\alpha.
\end{equation}
Using Eq.~\eqref{eqn:canonical}, Eqs.~\eqref{eqn:hamilton} can be written as
\begin{equation}
  (\ud H)_A (\varepsilon) + O(\varepsilon^2) = \Omega_{AB} \dot{\Gamma}^B (\varepsilon),
\end{equation}
where $\Gamma(\varepsilon)$ is the path that the particle takes through phase space, and is parameterized also by $\tau$.
Moreover, $\Omega_{AB}$ is non-degenerate and has an inverse, $\Omega^{AB}$, such that
\begin{equation} \label{eqn:Omega_inverse}
  \Omega^{AC} \Omega_{CB} = \delta^A{}_B,
\end{equation}
so we may finally reexpress the equations of motion in the covariant form:
\begin{equation} \label{eqn:covariant_hamilton}
  \dot{\Gamma}^A (\varepsilon) = \Omega^{AB} (\ud H)_B (\varepsilon) + O(\varepsilon^2).
\end{equation}

\subsubsection{Geometric construction for the Hamiltonian phase space}

In the below analysis, we are interested in the differences between covariant representations of phase-space quantities at different points in $\tau$.
From such differences, we construct a covariant description of the evolution of quasi-conserved worldline quantities for the perturbed Hamiltonian system.

Before performing a perturbative analysis, we first review a few notions from differential geometry.
First, we recall the definition of pullbacks and pushforwards.
Consider some map $\phi: \mc M \to \widetilde{\mc M}$, where $\mc M$ and $\widetilde{\mc M}$ are arbitrary manifolds.
We will use capital Latin letters to indicate indices on $\mc M$, and capital Latin indices with tildes to indicate indices on $\widetilde{\mc M}$.
For some scalar field $\tilde f$ on $\widetilde{\mc M}$, there is a scalar field on $\mc M$, denoted by $\phi_* \tilde f$, that is defined by
\begin{equation}
  \phi_* \tilde f \equiv \tilde f \circ \phi.
\end{equation}
The operation $\phi_*$ is called the \emph{pullback} (as it goes in the opposite direction of $\phi$, from scalar fields on $\widetilde{\mc M}$ to those on $\mc M$).
Similarly, for any vector field $v^A$ on $\mc M$, there is a corresponding vector field $(\phi^* v)^{\tilde A}$ on $\widetilde{\mc M}$ that is defined by
\begin{equation} \label{eqn:pushforward_def}
  (\phi^* v) (\tilde f) \equiv v(\tilde f \circ \phi),
\end{equation}
where we consider $v^A$ and $(\phi^* v)^{\tilde A}$ as a differential operators acting on scalar fields; see Chapter~2 of Wald~\cite{Wald1984} (as there is no risk of confusion here, we do not bold vector fields when treating them as differential operators).
This definition is linear, and so we can define a linear map $(\phi^*)^{\tilde A}{}_A$ by
\begin{equation} \label{eqn:pushforward_tensor}
  (\phi^* v)^{\tilde A} = (\phi^*)^{\tilde A}{}_A v^A.
\end{equation}
This mapping is known as the \emph{pushforward}, and in coordinates it can be written as
\begin{equation}
  (\phi^*)^{\tilde \aleph}{}_{\aleph} = \frac{\partial \phi^{\tilde \aleph}}{\partial X^\aleph},
\end{equation}
where $\phi^{\tilde \aleph}$ denotes the coordinates of $\phi(X)$.
By comparing Eq.~\eqref{eqn:pushforward_tensor} to the definition of the exterior derivative of scalar fields, namely that
\begin{equation}
  v^A (\ud f)_A \equiv v(f),
\end{equation}
we find that
\begin{equation} \label{eqn:d_pullback}
  (\ud \phi_* \tilde f)_A = (\phi^*)^{\tilde A}{}_A (\ud \tilde f)_{\tilde A}.
\end{equation}

Next, recall that, as phase space is a fiber bundle (the cotangent bundle), there is a natural projection $\pi: T^* \mc M \to \mc M$.
This projection can be used to generate a curve $\gamma(\varepsilon)$ that can be defined by
\begin{equation} \label{eqn:pi_Gamma}
  \gamma(\tau, \varepsilon) \equiv \pi[\Gamma(\tau, \varepsilon)],
\end{equation}
which is a curve in $\mc M$ that is determined by solutions to Hamilton's equations.

For our below exploration of flux-balance laws in Sec.~\ref{sec:flux_balance}, we are interested in the evolution of worldline quantities defined on phase space, and their relation to the field $\phi$ that is sourced by the worldline motion.
Because the field values are determined by the phase-space trajectory of the worldline motion, those field values themselves may be regarded as functions of phase space $\phi(x, X)$.
Such field values are the subclass of fields that can be sourced by a particular Hamiltonian system, and so have properties dependent on the source equation of motion.

In the below Sec.~\ref{sec:flux_balance}, we make use of the derivatives of field quantities with respect to phase space coordinates $X$, denoted by $\cal D_A$.
To derive the relationship of such quantities to worldline evolution, we then need to understand the relationship between such derivatives on phase space and the evolution of worldline quantities.
The remainder of this section is given to a formalization of the necessary phase space identities for our later derivation.

We next use the notion of pullbacks and pushforwards to define a propagator in phase space, in order to relate indices at different points.
Since Hamilton's equations are a set of first-order ordinary differential equations, there is a map $\flow(\Delta \tau, \varepsilon)$ that maps any point $X = \Gamma(\tau, \varepsilon)$ to a point $X' = \Gamma(\tau + \Delta \tau, \varepsilon)$.
We refer to the map $\flow$ as the \emph{Hamilton flow map}.
This is a map from phase space to itself, and so possesses a pushforward $(\flow^*)^{A'}{}_A (\Delta \tau, \varepsilon)$.

Note that this pushforward is a bitensor on phase space: it is a tensor field at two different points, $X$ and $X'$, which is defined so long as $X$ and $X'$ are both on a single curve $\Gamma(\varepsilon)$ through phase space, such that $X = \Gamma(\tau, \varepsilon)$ and $X' = \Gamma(\tau + \Delta \tau, \varepsilon)$.
In particular, it is \emph{only} defined at such pairs of points; if one takes a derivative of this bitensor (in some sense) with respect to $X'$, then the point $X$ must move in order for $\Delta \tau$ to stay fixed.
Instead of $(\flow^*)^{A'}{}_A (\Delta \tau, \varepsilon)$, we will therefore work with a subtly different bitensor, which we denote by $\flow^{A'}{}_A (\varepsilon)$ and call the \emph{Hamilton propagator}:
\begin{equation}
  \flow^{A'}{}_A (\varepsilon) \equiv (\flow^*)^{A'}{}_A [\Delta \tau(X', X; \varepsilon), \varepsilon],
\end{equation}
where $\Delta \tau(X', X; \varepsilon)$ is such that
\begin{equation}
  X = \Gamma(\tau, \varepsilon), \qquad X' = \Gamma[\tau + \Delta \tau(X', X; \varepsilon), \varepsilon].
\end{equation}
When one takes derivatives of the Hamilton propagator with respect to $X$ or $X'$, the other point does not move with it, as $\Delta \tau$ is no longer fixed.

In this paper, we use four key properties of the Hamilton propagator:
\begin{itemize}
\item Composition of the Hamilton propagator
\begin{equation} \label{eqn:hp_compose}
  \flow^A{}_{A'} (\varepsilon) \flow^{A'}{}_B (\varepsilon) = \delta^A{}_B.
\end{equation}
\item Derivative of the Hamilton flow map
\begin{equation} \label{eqn:d_hp}
  \{\ud [f \circ \flow(\tau' - \tau, \varepsilon)]\}_A = \left.\flow^{A'}{}_A (\varepsilon) (\ud f)_{A'}\right|_{X' = \Gamma(\tau', \varepsilon)},
\end{equation}
\item Equation of motion of the Hamilton propagator
\begin{equation} \label{eqn:lie_hp}
  \lie_{\dot \Gamma' (\varepsilon)} \flow^{A'}{}_A (\varepsilon) = 0.
\end{equation}
Note that this Lie derivative only acts at the point $X'$.
\item Hamilton propagation of the symplectic two-form
\begin{equation} \label{eqn:smorphism}
  \Omega_{AB} = \flow^{A'}{}_A (\varepsilon) \flow^{B'}{}_B (\varepsilon) \Omega_{A'B'} + O(\varepsilon^2).
\end{equation}
\end{itemize}

{\bf Composition of the Hamilton propagator:} The composition property follows from the composition identity for the Hamilton flow map:
\begin{equation}
  \flow(\Delta \tau_1, \varepsilon) \circ \flow(\Delta \tau_2, \varepsilon) = \flow(\Delta \tau_1 + \Delta \tau_2, \varepsilon).
\end{equation}
As such, applying the definition of the pushforward \eqref{eqn:pushforward_def}, we conclude that
\begin{equation}
  \flow^{A'}{}_{A''} (\varepsilon) \flow^{A''}{}_A (\varepsilon) = \flow^{A'}{}_A (\varepsilon).
\end{equation}
Similarly, since $\flow(0, \varepsilon)$ is the identity,
\begin{equation} \label{eqn:hp_inverse}
  \flow^A{}_{A'} (\varepsilon) \flow^{A'}{}_B (\varepsilon) = \delta^A{}_B.
\end{equation}

{\bf Derivative of the Hamilton flow map:} The derivative property comes directly from the derivative of the pullback [Eq.~\eqref{eqn:d_pullback}].
 Suppose that one has a scalar field $f$ that is evaluated at $X' = \Gamma(\tau', \varepsilon)$.
There is another scalar field, $f \circ \flow(\tau' - \tau, \varepsilon)$, which is evaluated at $X = \Gamma(\tau, \varepsilon)$.
So long as the relations between $X$ and $\tau$ and $X'$ and $\tau'$ remain fixed, it follows from Eq.~\eqref{eqn:d_pullback} that
\begin{equation}
  \{\ud [f \circ \flow(\tau' - \tau, \varepsilon)]\}_A = \left.\flow^{A'}{}_A (\varepsilon) (\ud f)_{A'}\right|_{X' = \Gamma(\tau', \varepsilon)},
\end{equation}
where (for brevity) we have dropped the explicit point that each side is being evaluated at $X = \Gamma(\tau, \varepsilon)$.

{\bf Equation of motion of the Hamilton propagator:} To determine the equation of motion, we use the definition of the Lie derivative from, for example, Eq. (C.2.1) Wald~\cite{Wald1984}, in terms of pushforwards; in our notation, where we note that $\flow(\Delta \tau, \varepsilon)$ \emph{is} the diffeomorphism that moves you along the integral curve of $\dot \Gamma^A$, we have that
\begin{widetext}
\begin{equation}
  \begin{split}
    \lie_{\dot \Gamma' (\varepsilon)} \flow^{A'}{}_A (\varepsilon) &\equiv \frac{\ud}{\ud \tau''} \bigg\{(\flow^*)^{A'}{}_{A''} (\tau'' - \tau', \varepsilon) \left.\flow^{A''}{}_A (\varepsilon)\right|_{X'' = \Gamma(\tau'', \varepsilon)}\bigg\}\bigg|_{\tau'' = \tau'} \\
    &= \left.\frac{\ud}{\ud \tau''} \flow^{A'}{}_A (\varepsilon)\right|_{\tau'' = \tau} = 0,
  \end{split}
\end{equation}
\end{widetext}
where we have simplified the argument of the derivative using the composition property of the Hamilton propagator [Eq.~\eqref{eqn:hp_compose}].

{\bf Hamilton propagation of the symplectic two-form:} One particular application of the equation of motion of the Hamilton propagator [Eq.~\eqref{eqn:lie_hp}] is to derive the Hamilton propagation of $\Omega_{AB}$.
As the techniques we use here are useful in Sec.~\ref{sec:hamilton_pert}, we go through this calculation in detail here.
To prove Eq.~\eqref{eqn:smorphism}, we first use Cartan's magic formula,
\begin{equation} \label{eqn:cartan}
  \lie_{\dot \Gamma(\varepsilon)} \Omega_{AB} = \dot \Gamma^C (\varepsilon) (\ud \Omega)_{CAB} + \{\ud [\dot \Gamma(\varepsilon) \cdot \Omega]\}_{AB}
\end{equation}
(where $\cdot$ denotes contraction of a vector with the first index of a differential form), and then use Eq.~\eqref{eqn:covariant_hamilton} to show that
\begin{equation}
  \dot{\bs \Gamma} (\varepsilon) \cdot \bs \Omega = \ud H (\varepsilon) + O(\varepsilon^2),
\end{equation}
and so the second term in Eq.~\eqref{eqn:cartan} is $O(\varepsilon^2)$, since $\ud^2 = 0$.
The symplectic two-form is closed [Eq~\eqref{eqn:d_Omega}], so the first term vanishes, and we find that
\begin{equation} \label{eqn:lie_Omega}
  \lie_{\dot \Gamma(\varepsilon)} \Omega_{AB} = O(\varepsilon^2).
\end{equation}
We can then solve this equation by considering the following expression:
\begin{equation} \label{eqn:d_tau_Omega}
  \begin{split}
    \frac{\ud}{\ud \tau'} \left[\flow^{A'}{}_A (\varepsilon) \flow^{B'}{}_B (\varepsilon) \Omega_{A'B'}\right] = O(\varepsilon^2),
  \end{split}
\end{equation}
which follows from the equation of motion of the Hamilton propagator [Eq.~\eqref{eqn:lie_hp}] and Eq.~\eqref{eqn:lie_Omega}, together with the fact that the argument of the derivative in Eq.~\eqref{eqn:d_tau_Omega} is now a scalar at $X'$, and for any scalar field $f$,
\begin{equation}
  \frac{\ud f}{\ud \tau} = \dot{\Gamma}^A (\ud f)_A = \lie_{\dot \Gamma} f.
\end{equation}
We have now converted the problem into an ordinary differential equation in $\tau'$, which can be solved using the initial condition that $\Omega_{A'B'}$ at $\tau' = \tau$ is just $\Omega_{AB}$:
\begin{equation}
  \flow^{A'}{}_A (\varepsilon) \flow^{B'}{}_B (\varepsilon) \Omega_{A'B'} = \Omega_{AB} + O(\varepsilon^2).
\end{equation}
Inverting the Hamilton propagators using the composition identity Eq.~\eqref{eqn:hp_inverse}, we recover the Hamilton propagation of the symplectic two-form Eq.~\eqref{eqn:smorphism}.

Starting instead from
\begin{equation}
  \lie_{\dot \Gamma} \Omega^{AB} = O(\varepsilon^2),
\end{equation}
which follows from the symplectic two-form equation of motion [Eq.~\eqref{eqn:lie_Omega}] and the definition of the inverse of the symplectic two-form [Eq.~\eqref{eqn:Omega_inverse}], a similar derivation shows that
\begin{equation} \label{eqn:smorphism_inverse}
  \Omega^{A'B'} = \flow^{A'}{}_A (\varepsilon) \flow^{B'}{}_B (\varepsilon) \Omega^{AB} + O(\varepsilon^2).
\end{equation}

\subsubsection{Perturbative analysis} \label{sec:hamilton_pert}

Unlike in the case of conserved quantities, the perturbative analysis is somewhat complicated by the fact that the quantity of interest, namely $\Gamma(\tau, \varepsilon)$, is not a scalar, but a function that returns points on phase space $T^* \mc M$.
Previously, we had considered this function at fixed $\varepsilon$, and varied $\tau$ to obtain a curve $\Gamma(\varepsilon)$.
However, one can instead fix $\tau$, and vary $\varepsilon$ to obtain a different curve in $T^* \mc M$; at $\Gamma(\tau)$, we denote the tangent vector to this curve by $\var \Gamma^A$.

The aim of the present derivation is to determine the tangent vector $\delta \Gamma^A$ as a function of $\tau$.
The vector $\delta \Gamma^A$ encodes the dependence of worldline phase space quantities on the field $\phi^R$ that perturbs the worldline motion.
Ultimately, we show that the difference of $\delta \Gamma^A$ at different points on the worldline is equivalent to an integral over the field-dependent forcing term of the Hamiltonian:
\begin{equation} \label{eqn:Delta_var_Gamma_preview}
  \flow^A{}_{A'} \var \Gamma^{A'} - \var \Gamma^A = \Omega^{AB} \int_\tau^{\tau'} \ud \tau'' \flow^{B''}{}_B (\ud \var H)_{B''}.
\end{equation}
This expression is the first step required for our derivation in Sec.~\ref{sec:integrated} of the flux-balance law associated with phase-space symmetry operators -- it establishes the relationship between differences in worldline phase-space quantities and the local field that appears in the Hamiltonian.
The remaining steps to relate the field values on the worldline to fluxes on a remote worldtube are given in Sec.~\ref{sec:integrated}.

To work towards proving Eq.~\eqref{eqn:Delta_var_Gamma_preview}, one would hope to obtain a differential equation by writing something like
\begin{equation}
  \frac{\uD}{\ud \tau} \var \Gamma^A = \ldots,
\end{equation}
and then integrate this equation.
However, the right-hand side, since it involves the derivative of a vector field, requires a connection on phase space---something which we do not possess.
We therefore need a notion of a derivative of a vector field along a curve that does not require a connection: this can be given by the Lie derivative.

To compute the Lie derivative of $\var \Gamma^A$ with respect to $\dot \Gamma^A$, we use the definition of vector fields as differential operators acting on scalar fields mentioned below Eq.~\eqref{eqn:pushforward_def}.
For the specific cases of the tangent vectors $\var \Gamma^A$, $\dot \Gamma^A(\epsilon)$, and $\dot \Gamma^A = \dot \Gamma^A(\epsilon)|_{\epsilon = 0}$, the differential operators are defined by
\begin{subequations}
\begin{align}
  \var \Gamma(f)\Big|_{\Gamma(\tau)} &\equiv \left.\frac{\partial}{\partial \varepsilon} f[\Gamma(\tau, \varepsilon)]\right|_{\varepsilon = 0} \label{eqn:delta_Gamma},\\
  \dot \Gamma (f, \varepsilon)\Big|_{\Gamma(\tau, \varepsilon)} &\equiv \left.\frac{\partial}{\partial \tau'} f[\Gamma(\tau', \varepsilon)]\right|_{\tau' = \tau} \label{eqn:dot_Gamma_of_epsilon},
\end{align}
\end{subequations}
for any scalar field on phase space $f$.
Then,
\begin{equation}
  \dot \Gamma(f)\Big|_{\Gamma(\tau)} = \left.\dot \Gamma(f, \epsilon)\Big|_{\Gamma(\tau, \epsilon)} \right|_{\epsilon = 0} = \left.\frac{\partial}{\partial \tau'} f[\Gamma(\tau')]\right|_{\tau' = \tau} \label{eqn:dot_Gamma}.
\end{equation}
In these equations, we make explicit that the vector fields are at $\Gamma(\tau)$.

The Lie derivative of $\var \Gamma^A$ with respect to $\dot \Gamma^A$ is then given by the commutator \{see Eqs.~(C.2.7) and~(2.2.14) of Wald~\cite{Wald1984}\}:
\begin{equation} \label{eqn:lie_Gamma_commutator}
  (\lie_{\dot \Gamma} \var \Gamma) (f)\Big|_{\Gamma(\tau)} \equiv \dot \Gamma [\var \Gamma(f)]\Big|_{\Gamma(\tau)} - \var \Gamma [\dot \Gamma(f)]\Big|_{\Gamma(\tau)}.
\end{equation}
The first of the expressions on the right-hand side is easy to compute:
\begin{equation} \label{eqn:dot_delta_Gamma}
  \begin{split}
    \dot \Gamma [\var \Gamma(f)]\Big|_{\Gamma(\tau)} &= \left.\frac{\partial}{\partial \tau'} \var \Gamma(f)\Big|_{\Gamma(\tau')}\right|_{\tau' = \tau} \\
    &= \left.\frac{\partial^2}{\partial \tau' \partial \varepsilon} f[\Gamma(\tau', \varepsilon)]\right|_{\tau' = \tau,\, \varepsilon = 0} \\
    &= \frac{\partial}{\partial \epsilon} \left.\dot \Gamma (f, \epsilon) \big|_{\Gamma (\tau, \epsilon)} \right|_{\epsilon = 0},
  \end{split}
\end{equation}
where we have first expanded the tangent vectors using Eqs.~\eqref{eqn:dot_Gamma} and~\eqref{eqn:delta_Gamma}, then, using the commutativity of the partial derivatives simplified the $\epsilon$-dependent field expression using Eq.~\eqref{eqn:dot_Gamma_of_epsilon}.

The second expression on the right-hand side of the commutator form of the Lie derivative~\eqref{eqn:lie_Gamma_commutator}, on the other hand, is given by
\begin{equation} \label{eqn:delta_dot_Gamma}
  \var \Gamma [\dot \Gamma(f)]\Big|_{\Gamma(\tau)} = \left.\frac{\partial}{\partial \varepsilon} \dot \Gamma(f)\Big|_{\Gamma(\tau, \varepsilon)}\right|_{\varepsilon = 0}.
\end{equation}
Note that this is \emph{not} the same as the right-hand side of the second equality of Eq.~\eqref{eqn:dot_delta_Gamma} -- in Eq.~\eqref{eqn:dot_delta_Gamma}, we include the $\epsilon$ variation of the tangent vector as well as the worldline point at which the scalar is evaluated, where in Eq.~\eqref{eqn:delta_dot_Gamma}, the $\epsilon$ dependence is confined to only the worldline point.

As such, we have that
\begin{equation} \label{eqn:lie_dot_var_inter}
  (\lie_{\dot \Gamma} \var \Gamma) (f)\Big|_{\Gamma(\tau)} = \left.\frac{\partial}{\partial \epsilon} \left[\dot \Gamma (f, \varepsilon)\Big|_{\Gamma(\tau, \varepsilon)} - \dot \Gamma(f)\Big|_{\Gamma(\tau, \varepsilon)}\right]\right|_{\varepsilon = 0}.
\end{equation}
Using the covariant form of the equations of motion [Eq.~\eqref{eqn:covariant_hamilton}], we find that the right-hand side can be written in terms of the Hamiltonian $H(\varepsilon)$:
\begin{equation}
  \begin{split}
    \dot \Gamma (f, \varepsilon) - \dot \Gamma(f) &= \Omega^{AB} (\ud f)_A \left\{\ud \left[H(\varepsilon) - H\right]\right\}_B + O(\varepsilon^2) \\
    &= \varepsilon \Omega^{AB} (\ud f)_A (\ud \var H)_B + O(\varepsilon^2),
  \end{split}
\end{equation}
where this equation is evaluated at $\Gamma(\tau, \varepsilon)$.
Evaluating the derivative in~\eqref{eqn:lie_dot_var_inter} therefore yields
\begin{equation}
  (\lie_{\dot \Gamma} \var \Gamma) (f)\Big|_{\Gamma(\tau)} = \Omega^{AB} (\ud f)_A (\ud \var H)_B\Big|_{\Gamma(\tau)},
\end{equation}
and so
\begin{equation} \label{eqn:lie_dot_var}
  \lie_{\dot \Gamma} \var \Gamma^A = \Omega^{AB} (\ud \var H)_B.
\end{equation}

To solve Eq.~\eqref{eqn:lie_dot_var}, we can use similar logic as was used to determine the Hamilton propagation of the symplectic two-form [Eq.~\eqref{eqn:smorphism}] from its Lie derivative with respect to $\Dot \Gamma$ [Eq.~\eqref{eqn:cartan}].
We use the equation of motion of the Hamilton propagator [Eq.~\eqref{eqn:lie_hp}] and the Hamilton propagation of the inverse symplectic form [Eq.~\eqref{eqn:smorphism_inverse}] to re-express Eq.~\eqref{eqn:lie_dot_var} as a Lie derivative of a quantity that is a scalar at $\Gamma(\tau, \epsilon)$ and a vector at $\Gamma(\tau', \epsilon)$.
Then, the Lie derivative may be replaced with $\ud/\ud\tau'$ and we have that
\begin{equation}
  \frac{\ud}{\ud \tau'} \left(\flow^A{}_{A'} \var \Gamma^{A'}\right) = \Omega^{AB} \flow^{B'}{}_B (\ud \var H)_{B'}.
\end{equation}
This ordinary differential equation can be integrated, yielding
\begin{equation} \label{eqn:Delta_var_Gamma}
  \flow^A{}_{A'} \var \Gamma^{A'} - \var \Gamma^A = \Omega^{AB} \int_\tau^{\tau'} \ud \tau'' \flow^{B''}{}_B (\ud \var H)_{B''},
\end{equation}
where $X'' = \Gamma(\tau'')$.
Typically, we will assume that $\var \Gamma^A = 0$, and so we will only concern ourselves with the right-hand side.

\subsubsection{Action-angle variables}

We next assume that the worldline motion determined by the background Hamiltonian $H$ is completely integrable, with $n$ constants of motion $P_\alpha$ satisfying
\begin{gather}
  \frac{\ud P_\alpha}{\ud \tau} = \Omega^{AB} (\ud P_\alpha)_A (\ud H)_B = 0, \\
  \Omega^{AB} (\ud P_\alpha)_A (\ud P_\beta)_B = 0,
\end{gather}
and where the $(\ud P_\alpha)_A$ are linearly independent (the indices $\alpha$ and $\beta$ range from $0$ to $n - 1$).
We are primarily interested in states of the system in the neighborhood of level sets $\mathcal M_{\bs P}$ of these constants of motion:
\begin{equation}
  \mathcal M_{\bs P} = \{X \in T^* \mathcal M \mid P_\alpha (X) = P_\alpha\}.
\end{equation}

In the case of interest, bound motion in the Kerr spacetime, a generalization to the Liouville-Arnold Theorem~\cite{Arnold1989, Fiorani2003, Hinderer2008} shows that there exist coordinates in a neighborhood of $\mathcal M_{\bs P}$, known as \emph{action-angle variables}
\begin{equation}
  X^\aleph \equiv \begin{pmatrix}
    q^\alpha \\
    J_\alpha
  \end{pmatrix},
\end{equation}
such that
\begin{itemize}

\item the $J_\alpha$ (the \emph{action} variables) are constants of motion, where each $P_\alpha$ is a function only of $J_\alpha$;

\item the $q^\alpha$ (the \emph{angle} variables) are such that $q^0$ is a non-compact coordinate, while $q^1, \cdots, q^n$ are periodic in $2\pi$; and

\item this set of coordinates is \emph{symplectic}; that is,
\begin{equation} \label{eqn:symplectic_coords}
  \bs \Omega = \ud J_\alpha \wedge \ud q^\alpha.
\end{equation}

\end{itemize}

The main utility of these coordinates is that the Hamilton propagator takes a particularly simple form in these coordinates.
This follows from the fact that Hamilton's equations take the form
\begin{subequations}
  \begin{align}
    \frac{\ud J_\alpha}{\ud \tau} &= 0, \\
    \frac{\ud q^\alpha}{\ud \tau} &\equiv \nu^\alpha (\bs J).
  \end{align}
\end{subequations}
Using the fact that $\flow^{\aleph'}{}_\aleph = \partial X^{\aleph'}/\partial X^\aleph$, it follows that
\begin{equation}
  \begin{split}
    \flow^{A'}{}_A &= (\partial_{q^\alpha})^{A'} \left[(\ud q^\alpha)_A + (\tau' - \tau) \frac{\partial \nu^\alpha}{\partial J_\beta} (\ud J_\beta)_A\right] \\
    &\hspace{1em}+ (\partial_{J_\alpha})^{A'} (\ud J_\alpha)_A.
  \end{split}
\end{equation}

These coordinates are also useful because the constants of motion can be determined entirely from the action variables.
Given changes $\Delta J_\alpha$ in the action variables due to the perturbing scalar field, the corresponding changes in the constants of motion can be determined entirely from the relationship $P_\alpha (\bs J)$.
In particular, this allows one to determine changes in the usual constants of motion considered for bound orbits in the Kerr spacetime: $m^2 \equiv -Q_g$, $E \equiv -E_{\partial_t}$, $L_z \equiv E_{\partial_\phi}$, and $Q \equiv Q_K$, where $K_{ab}$ is the usual Carter Killing tensor in Kerr:
\begin{equation}
  K_{ab} = r^2 g_{ab} + \frac{1}{\Delta} v^+_{(a} v^-_{b)},
\end{equation}
where
\begin{equation}
  (v^\pm)^a \equiv (r^2 + a^2) (\partial_t)^a + a (\partial_\phi)^a \pm \Delta (\partial_r)^a
\end{equation}
and $\Delta \equiv r^2 - 2M r + a^2$.
Note, however, that $P_\alpha (\bs J)$ is not given by a known, closed-form expression; as such, determining the evolution of the Carter constant $Q$ (for example) requires both a specification of the action variables and a numerical inversion of the (known) expressions for the function $J_\alpha (\bs P)$ in~\cite{Hinderer2008}, which is one-to-one in a neighborhood of the submanifold of constant $P_\alpha$.

\section{Flux-balance laws} \label{sec:flux_balance}

\subsection{Conserved currents}

Consider a theory for a field $\Phi_{\ms A}$, where capital, script Latin indices indicate some collection of indices associated with the field (in the case where $\Phi_{\ms A}$ denotes the vector potential, $\ms A = a$, while if $\Phi_{\ms A}$ denotes the metric, $\ms A = ab$, etc.).
We denote by $\bs L$ the \emph{Lagrangian four-form}, which is a functional of $\Phi_{\ms A}$.
For brevity, we do not denote the dependence in such functionals on $\Phi_{\ms A}$ explicitly.
The utility of considering the Lagrangian four-form, instead of the action (its integral), is provided by the fact that one does not need to worry about whether any of the integrals that arise are finite.

The equations of motion arise from a variation of the Lagrangian four-form by
\begin{equation} \label{eqn:var_L}
  \var \bs L = \bs E^{\ms A} \var \Phi_{\ms A} + \ud \bs \theta\{\var \bs \Phi\}.
\end{equation}
The term $\bs E^{\ms A}$ in the first term is a functional of $\Phi_{\ms A}$, and reflects the equations of motion: for a free field theory, the equations of motion read
\begin{equation}
  \bs E^{\ms A} = 0.
\end{equation}
The second term, the exterior derivative of the \emph{presymplectic form} $\bs \theta\{\var \bs \Phi\}$, defines the presymplectic form up to a closed three-form.
This three-form is a \emph{linear} functional of $\var \Phi_{\ms A}$; this we indicate explicitly using curly brackets (in the rare cases where we will need to denote the dependence of a nonlinear functional, we will use the traditional square brackets).
From this presymplectic form, one can define the \emph{symplectic current} from two variations, $\var_1$ and $\var_2$:
\begin{equation} \label{eqn:omega}
  \begin{split}
    \bs \omega\{\var_1 \bs \Phi, \var_2 \bs \Phi\} &\equiv \var_1 \bs \theta\{\var_2 \bs \Phi\} - \bs \theta\{\var_1 \var_2 \bs \Phi\}\\
    &\hspace{1em}- (\interchange{\var_1}{\var_2}).
  \end{split}
\end{equation}
This three-form current is bilinear and antisymmetric in $\var_1 \Phi_{\ms A}$ and $\var_2 \Phi_{\ms A}$, and is moreover independent of $\var_1 \var_2 \Phi_{\ms A}$, even if the two variations are not independent.

In this paper, we will not need any properties of the symplectic current other than the fact that Eqs.~\eqref{eqn:var_L} and~\eqref{eqn:omega} imply that, assuming that $\var_{1, 2}$ and $\ud$ commute,
\begin{equation} \label{eqn:domega}
  \ud \bs \omega\{\var_1 \bs \Phi, \var_2 \bs \Phi\} = \var_1 \Phi_{\ms A} \ord{1}{\bs E}^{\ms A \ms B} \var_2 \Phi_{\ms B} - (\interchange{\var_1}{\var_2}),
\end{equation}
where the linear operator $\ord{1}{\bs E}^{\ms A \ms B}$ is defined by
\begin{equation}
  \var \bs E^{\ms A} \equiv \ord{1}{\bs E}^{\ms A \ms B} \var \Phi_{\ms B}.
\end{equation}
This implies that the symplectic current is conserved, provided that the linearized equations of motion hold for $\var_1 \Phi_{\ms A}$ and $\var_2 \Phi_{\ms A}$:
\begin{equation}
  \ord{1}{\bs E}^{\ms A \ms B} \var_1 \Phi_{\ms B} = \ord{1}{\bs E}^{\ms A \ms B} \var_2 \Phi_{\ms B} = 0.
\end{equation}
Since Eq.~\eqref{eqn:domega} is the only feature of the symplectic current which we use, the discussion in the rest of this paper holds for any other bilinear current defined on the space of variations which differs from the symplectic current by a closed three-form.

In particular, given any linear differential operator $\bs{\mc O}^{\ms A \ms B}$, there exists a bilinear current $\bs J_{\mc O}$ and an operator, the adjoint $(\bs{\mc O}^\dagger)^{\ms A \ms B}$, such that~\cite{Wald1978}
\begin{equation}
  \ud \bs J_{\mc O} \{\bs \Phi, \bs \Psi\} \equiv \Phi_{\ms A} \bs{\mc O}^{\ms A \ms B} \Psi_{\ms B} - \Psi_{\ms A} (\bs{\mc O}^\dagger)^{\ms A \ms B} \Phi_{\ms B}.
\end{equation}
The current $\bs J_{\mc O}$ is unique up to a closed three-form.
Equation~\eqref{eqn:domega} shows that the operator $\ord{1}{\bs E}^{\ms A \ms B}$ is self-adjoint, with a choice of this current $\bs J_{\ord{1} E}$ being the symplectic current $\bs \omega$.
Some references (see, for example,~\cite{Torres_del_Castillo1999}) use the self-adjointness of $\ord{1}{\bs E}^{\ms A \ms B}$ as the \emph{motivation} for the construction of this current, whereas the approach presented here constructs this current explicitly from the Lagrangian.

In order to derive a flux-balance law, one needs a current that is conserved in the absence of sources.
As described above, such a current arises in the form of the symplectic current, which depends on two solutions to the linearized field equations.
However, one typically only has one such solution $\var \Phi_{\ms A}$; in order to form a nonzero conserved current from $\var \Phi_{\ms A}$ alone, we need a mapping from the space of solutions to the linearized field equations to itself.
Such a mapping is called a \emph{symmetry operator}.

Explicitly, we define a symmetry operator $\mc D_{\ms A}{}^{\ms B}$ as a linear operator acting on $\var \Phi_{\ms A}$ such that
\begin{equation} \label{eqn:symmetry_op}
  \ord{1}{\bs E}^{\ms A \ms C} \mc D_{\ms C}{}^{\ms B} = \widetilde{\mc D}^{\ms A}{}_{\ms C} \ord{1}{\bs E}^{\ms C \ms B},
\end{equation}
for some other operator $\widetilde{\mc D}^{\ms A}{}_{\ms B}$.
In terms of these operators, we have that
\begin{equation} \label{eqn:d_symm}
  \begin{split}
    \ud \bs \omega\{\var_1 \bs \Phi, \bs{\mc D} \cdot \var_2 \bs \Phi\} &= \var_1 \Phi_{\ms A} \widetilde{\mc D}^{\ms A}{}_{\ms C} \ord{1}{\bs E}^{\ms C \ms B} \var_2 \Phi_{\ms B} \\
    &\hspace{1em}- (\mc D_{\ms A}{}^{\ms C} \var_2 \Phi_{\ms C}) \ord{1}{\bs E}^{\ms A \ms B} \var_1 \Phi_{\ms B}.
  \end{split}
\end{equation}
Here, we still consider two different linearized fields $\var_1 \Phi_{\ms A}$ and $\var_2 \Phi_{\ms A}$, since in the case of interest we do have two such fields, although they will not both be solutions to the free, linearized equations of motion---there will be source terms.

This discussion so far has been applicable to general theories that can be constructed from a Lagrangian.
In this paper, we specialize to the case of a massless scalar field $\phi$, with a Lagrangian four-form
\begin{equation}
  \bs L = \frac{1}{2} \bs \epsilon (\nabla^a \phi) (\nabla_a \phi).
\end{equation}
It then follows that
\begin{equation}
  \bs E = -\bs \epsilon \Box \phi, \qquad \theta_{abc} \{\var \phi\} = (\var \phi \nabla^d \phi) \epsilon_{dabc},
\end{equation}
where the latter follows from the fact that \{see, for example, Eq.~(B.2.22) of Wald~\cite{Wald1984}\}
\begin{equation}
  (\nabla_e v^e) \bs \epsilon = \ud (\bs v \cdot \bs \epsilon).
\end{equation}
As such, we have that
\begin{align}
  \ord{1}{\bs E} &= -\bs \epsilon \Box, \\
  \omega_{abc} \{\var_1 \phi, \var_2 \phi\} &= \epsilon_{dabc} \left[\var_2 \phi \nabla^d \var_1 \phi - (\interchange{\var_1}{\var_2})\right].
\end{align}
Furthermore, since the Lagrangian is quadratic in $\phi$, and so the equations of motion linear in $\phi$, $\phi$ itself can be considered a variation by defining a one-parameter family of scalar fields by $\phi(\varepsilon) \equiv \varepsilon \phi$.
In this paper, we take advantage of this fact by dropping the variation symbols in the arguments of the symplectic current, writing $\phi_1$ and $\phi_2$ instead of $\var_1 \phi$ and $\var_2 \phi$.

\subsection{Integrated flux-balance laws} \label{sec:integrated}

In Sec.~\ref{sec:eom}, we wrote the quantities of interest, such as first-order changes in conserved quantities $\var \Delta E_\xi (\tau', \tau)$, as integrals over the worldline: see the earlier expressions for the change in quantities associated with Killing vectors $E_\xi$ [Eq.~\eqref{eqn:Delta_E_xi}], those associated with Killing tensors $Q_K$ [Eq.~\eqref{eqn:Delta_Q_K}], and general perturbations of $\Gamma(\tau, \epsilon)$ [Eq.~\eqref{eqn:Delta_var_Gamma}].
In this section, we relate these integrals over the worldline to fluxes of conserved currents that we construct using the tools of the previous section.

\subsubsection{Spacetime symmetries}

One particular example of a symmetry operator is given by the Lie derivative with respect to a vector field: if $\lie_\xi g_{ab} = 0$, then
\begin{equation}
  \lie_\xi \Box - \Box \lie_\xi = 0.
\end{equation}
This is Eq.~\eqref{eqn:symmetry_op}, with $\mc D = \widetilde{\mc D} = \lie_\xi$.
As such, we can consider the symplectic current constructed by
\begin{equation}
  \bs{\mc E}_\xi \{\phi_1, \phi_2\} \equiv \bs \omega\{\phi_1, \lie_\xi \phi_2\},
\end{equation}
which is known as the \emph{canonical current} (see, for example,~\cite{Hollands2012, Bonga2019}).
This current is conserved if $\Box \phi_1 = \Box \phi_2 = 0$; in general, we have from Eq.~\eqref{eqn:d_symm} that
\begin{equation} \label{eqn:d_canonical}
  \ud \bs{\mc E}_\xi \{\phi_1, \phi_2\} = -\bs \epsilon [\phi_1 \lie_\xi \Box \phi_2 - (\lie_\xi \phi_2) \Box \phi_1].
\end{equation}
The analogue of this current in the gravitational case, up to a boundary term, is equivalent to the conserved current coming from the effective stress-energy tensor and the Killing vector $\xi^a$ (see~\cite{Hollands2012} for the gravitational case, \cite{Grant2022} for more general gravitational theories, and \cite{Bonga2019} for a similar result in the electromagnetic case).

We now consider the canonical current constructed by using $\phi_1 = \phi^+$ (the retarded field) and $\phi_2 = \phi^{\rm R}$ (the regular field), where these two scalar fields were defined by Eqs.~\eqref{eqn:ret_integral} and~\eqref{eqn:reg_integral}, respectively.
Using the field equations for the retarded and regular fields [Eqs.~\eqref{eqn:ret_eom} and \eqref{eqn:reg_eom}] with Eq.~\eqref{eqn:d_canonical}, we find that
\begin{equation}
  \ud \bs{\mc E}_\xi \{\phi^+, \phi^{\rm R}\} = \bs \epsilon \rho \lie_\xi \phi^{\rm R}.
\end{equation}
Integrating this equation over $\V{\tau', \tau}$ and using Stokes' theorem, the right-hand side becomes the negative of the right-hand side of our earlier equation for the change in $E_\xi$, and so Eq.~\eqref{eqn:Delta_E_xi} may be written as
\begin{equation} \label{eqn:E_xi_int_flux}
  \var \Delta E_\xi (\tau', \tau) = -\int_{\partial \V{\tau', \tau}} \bs{\mc E}_\xi \{\phi^+, \phi^{\rm R}\}.
\end{equation}
This equation is our flux-balance law: it relates quantities on the worldline to an integral of a conserved current.

\subsubsection{Second-order ``hidden symmetries''}

We now briefly consider the case of the Carter constant, which is generated not by an isometry of spacetime, but the existence of a Killing tensor.
First, note that there exists a symmetry operator, $\mc D_K$, which exists in the presence of a Killing tensor~\cite{Carter1977}:
\begin{equation}
  \mc D_K \phi \equiv \nabla_a (K^{ab} \nabla_b \phi).
\end{equation}
As such, one can define a conserved current analogous to the canonical current by
\begin{equation}
  \bs{\mc Q}_K \{\phi_1, \phi_2\} \equiv \bs \omega\{\phi_1, \mc D_K \phi_2\}.
\end{equation}
In Eq.~\eqref{eqn:Delta_Q_K}, we wrote down the formula for $\var \Delta Q_K$, giving the first-order change in the Carter constant.
One might hope that a flux balance law analogous to Eq.~\eqref{eqn:E_xi_int_flux} would hold, but one instead has that
\begin{equation}
  \begin{split}
    -2 \int_{\partial \V{\tau', \tau}} \bs{\mc Q}_K \{\phi^+, \phi^{\rm R}\} &= -2 \int \bs \epsilon \rho \nabla_a (K^{ab} \nabla_b \phi) \\
    &\neq \var \Delta Q_K.
  \end{split}
\end{equation}
As such, $\bs{\mc Q}_K \{\phi^+, \phi^{\rm R}\}$ seems to provide \emph{some} information about the worldline of the particle, but it is not clearly related to changes in the Carter constant.
Determining exactly what information is provided by this conserved current (as well as by generalizations to other field theories~\cite{Grant2019, Grant2020}) is outside of the scope of this paper, and will be pursued in future work.

Note that the fact that this conserved current is not directly applicable to the evolution of the Carter constant is suggested by the following property of the fluxes derived in~\cite{Mino2003}: while the flux-balance laws for $E$ and $L_z$ contain only information about the field at the horizon and infinity, the ``flux-balance laws'' for the Carter constant involve quantities averaged over the worldline of the particle.
This suggests that there is something fundamentally different about the Carter constant, and motivates considering the Hamiltonian approach, where all of the conserved quantities are on equal footing in terms of action variables.

\subsubsection{Hamiltonian systems}

We now show how to write the change in perturbations $\delta\Gamma^A$ given by Eq.~\eqref{eqn:Delta_var_Gamma} in the form of a flux-balance law.
A key realization about the type of symmetries discussed in this section is that they are entirely specialized to the system of Hamiltonian motion coupled to a field.
This is a notable departure from the symmetries discussed in the first two parts of this section, which hold for arbitrary field solutions $\phi$.

To derive the flux balance law associated with Hamiltonian worldline quantities, we first note that the retarded field $\phi^+$ can be considered as a function of the phase space coordinates $X$ of the worldline at $\tau$: write
\begin{equation}
  \phi^+ (x', X) = \int_{\V{\infty, -\infty}} \bs \epsilon'' G^+ (x', x'') \rho(x'', X),
\end{equation}
where, recalling that the worldline $\gamma(\tau'')$ may be written in terms of the projection of phase space points [Eq.~\eqref{eqn:pi_Gamma}], and that the Hamilton flow map $\Upsilon$ is used to map between different phase space points on the worldline,
\begin{equation}
  \rho(x', X) = \int_{-\infty}^\infty \ud \tau'' \delta\{x', [\pi \circ \flow(\tau'' - \tau)] (X)\}.
\end{equation}
This way of writing $\phi^+$ is inspired by similar techniques that arise in the two-timescale formulation of the self-force~\cite{two_timescale20XX}.
The key insight here is that one can now consider derivatives of $\phi^+$ and $\rho$ with respect to $X$; we will denote such derivatives by $\mc D_A$.
Since the retarded Green's function doesn't depend on the worldline, it follows that
\begin{equation}
  \Box \mc D_A \phi^+ = \mc D_A \rho,
\end{equation}
and so $\Box$ and $\mc D_A$ commute.

We now consider the following symplectic current:
\begin{equation}
  \bs{\mc J}_A^+ \{\phi\} \equiv \bs \omega\{\phi, \mc D_A \phi^+\},
\end{equation}
for any scalar field $\phi$, and where we evaluate these fields at some $x'$.
It follows that
\begin{equation}
  \begin{split}
    \left.\ud \bs{\mc J}_A^+ \{\phi\}\right|_{x'} = -\bs \epsilon' \{&\phi (x') \mc D_A \rho(x', X) \\
    &- [\mc D_A \phi^+ (x', X)] \Box' \phi(x')\}.
  \end{split}
\end{equation}
In the case where $\phi = \phi^{\rm R}$, the fact that $\phi^{\rm R}$ is source-free implies that
\begin{equation} \label{eqn:dJ_integrand}
  \left.\ud \bs{\mc J}_A^+ \{\phi^{\rm R}\}\right|_{x'} = -\bs \epsilon' \phi^{\rm R} (x') \mc D_A \rho(x', X).
\end{equation}
To evaluate $\mc D_A \rho(x', X)$, we use the derivative of the Hamilton flow map [Eq.~\eqref{eqn:d_hp}], so that
\begin{widetext}
\begin{equation}
  \mc D_A \rho(x', X) = q \int_{-\infty}^\infty \ud \tau'' \flow^{A''}{}_A \nabla_{A''} \delta[x', \pi(X'')]\Big|_{X'' = \flow(\tau'' - \tau) (X)}.
\end{equation}
At this point, we evaluate Eq.~\eqref{eqn:dJ_integrand} at some $x''$ and integrate over the volume $\V{\tau', \tau}$, yielding
\begin{equation} \label{eqn:J_dphi}
  \begin{split}
    \int_{\partial \V{\tau', \tau}} \bs{\mc J}_A^+ \{\phi^{\rm R}\} &= -q \int_{\V{\tau', \tau}} \bs \epsilon'' \phi^{\rm R} (x'') \int_{-\infty}^\infty \ud \tau''' \flow^{A'''}{}_A \nabla_{A'''} \delta[x'', \pi(X''')]\Big|_{X''' = \flow(\tau''' - \tau) (X)} \\
    &= -q \int_\tau^{\tau'} \ud \tau''' \flow^{A'''}{}_A \nabla_{A'''} \int_{\V{\infty, -\infty}} \bs \epsilon'' \phi^{\rm R} (x'') \delta[x'', \pi(X''')]\Big|_{X''' = \flow(\tau''' - \tau) (X)} \\
    &= -q \int_\tau^{\tau'} \ud \tau'' \flow^{A''}{}_A (\ud \phi^{\rm R})_{A''},
  \end{split}
\end{equation}
\end{widetext}
where in the second equality, we have switched the order of integration and updated the bounds according to Eq.~\eqref{eqn:delta_switch}, and in the third equality we have simply integrated over the delta function.
For brevity, we implicitly write $X'' \equiv \Gamma(\tau'') = \flow(\tau'' - \tau) (X)$.

Using the fact that $\var H = q \phi^{\rm R}$ [Eq.~\eqref{eqn:hamiltonian}], we now find that the integral on the right-hand side of Eq.~\eqref{eqn:J_dphi} is equivalent to the integral over the worldline we found in the right-hand side of our earlier result for the change in $\var \Gamma^A$ [Eq.~\eqref{eqn:Delta_var_Gamma}].
Combining these two equations, we therefore find that
\begin{equation} \label{eqn:Gamma_flux_balance}
  \flow^A{}_{A'} \var \Gamma^{A'} - \var \Gamma^A = -\Omega^{AB} \int_{\partial \V{\tau', \tau}} \bs{\mc J}_B^+ \{\phi^{\rm R}\}.
\end{equation}

This is of the form of a flux-balance law: it relates a ``change'' in the vector $\var \Gamma^A$ (defined in an appropriate way) to the integral of a conserved current.
Taking the ``action-variable'' component of Eq.~\eqref{eqn:Gamma_flux_balance}, and using the fact that
\begin{equation}
  \Omega^{AB} = 2 (\partial_{q^\alpha})^{[A} (\partial_{J_\alpha})^{B]}
\end{equation}
[which follows from Eq.~\eqref{eqn:symplectic_coords}], one finds the following formula in terms of coordinates:
\begin{equation}
  (\ud J_\alpha)_{A'} \var \Gamma^{A'} - (\ud J_\alpha)_A \var \Gamma^A = (\partial_{q^\alpha})^A \int_{\partial \V{\tau', \tau}} \bs{\mc J}_A^+ \{\phi^{\rm R}\}.
\end{equation}
Moreover, in coordinates, we have that (at any time $\tau$)
\begin{equation}
  J_\alpha (\tau, \varepsilon) = J_\alpha (\tau) + \varepsilon (\ud J_\alpha)_A \var \Gamma^A + O(\varepsilon^2),
\end{equation}
and so
\begin{equation} \label{eqn:J_flux_balance}
  \begin{split}
    \Delta J_\alpha (\tau', \tau) &\equiv J_\alpha (\tau') - J_\alpha (\tau) \\
    &= \varepsilon \int_{\partial \V{\tau', \tau}} \bs \omega\{\phi^{\rm R}, (\partial_{q^\alpha})^A \mc D_A \phi^+\} + O(\varepsilon^2).
  \end{split}
\end{equation}

\subsection{Averaged flux-balance laws} \label{sec:averaged}

Another type of flux-balance law that one can consider are those that are in an ``average'' form.
These flux-balance laws are useful in determining the average evolution of the quantities of interest, which is what one would expect a flux-balance law to physically provide.
The integrated flux-balance laws described in the previous subsection, while describing the full evolution of the system, do so in terms of a flux that is difficult to compute.
This is because the surface $\V{\tau, \tau'}$ which appears in these flux-balance laws intersects the worldline.
In contrast, the averaged flux-balance laws can be written in terms of a flux that is truly ``far away'' from the worldline (up to a caveat which we will discuss at the end of this section).

\begin{figure}
  \includegraphics[width=0.8\linewidth]{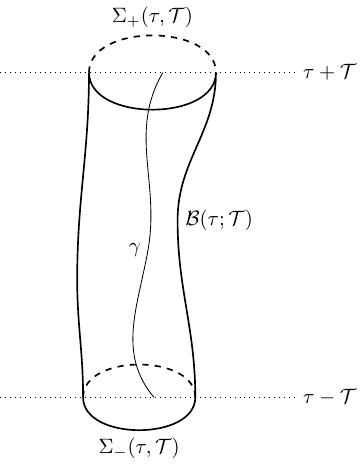}
  \caption{\label{fig:worldtube} The boundary of a region $\V{\tau - \mc T, \tau + \mc T}$ that surrounds the worldline $\gamma$ and intersects $\gamma$ at $\tau \pm \mc T$.
    This boundary is composed of three pieces: a worldtube $\B{\tau; \mc T}$ that surrounds $\gamma$ but does not intersect it, and two ``end-caps'' $\Sigma_\pm (\tau; \mc T)$.}
\end{figure}

To describe these averaged flux-balance laws, we consider the region $\V{\tau + \mc T, \tau - \mc T}$, where $\mc T$ is a quantity which will be taking to infinity.
Moreover, we split up the boundary of this region into three surfaces: a worldtube $\B{\tau; \mc T}$ which does not intersect $\gamma$, and two ``end-caps'' $\Sigma_\pm (\tau; \mc T)$ which intersect $\gamma$ at $\gamma(\tau \pm \mc T)$, respectively.
For simplicity, we assume that $\Sigma_\pm (\tau; \mc T)$ intersect $\gamma$ orthogonally, and that $\Sigma_\pm (\tau; \mc T)$ maintain the same rough size as $\mc T \to \infty$, and in particular do not become infinitely large in this limit.
These surfaces are shown in Fig.~\ref{fig:worldtube}.

We start with the case of flux-balance laws related to isometries, as those are simpler.
We first define
\begin{equation}
  \left\<\frac{\ud E_\xi}{\ud \tau}\right\> \equiv \lim_{\mc T \to \infty} \frac{\Delta E_\xi (\tau + \mc T, \tau - \mc T)}{2 \mc T}.
\end{equation}
From Eq.~\eqref{eqn:E_xi_int_flux}, we therefore have that
\begin{equation} \label{eqn:E_flux_balance_avg}
  \left\<\frac{\ud E_\xi}{\ud \tau}\right\> = -\lim_{\mc T \to \infty} \frac{\varepsilon}{2 \mc T} \int_{\B{\tau; \mc T}} \bs{\mc E}_\xi \{\phi^+, \phi^{\rm R}\} + O(\varepsilon^2).
\end{equation}
Here, we have dropped the contribution to the integral from the integrals over $\Sigmacap{\pm}{\tau; \mc T}$, since these terms stay finite in the limit $\mc T \to \infty$, and get divided by $\mc T$.\footnote{Some care must be taken here, as it is not clear that these integrals are even well-defined, due to the singularity of $\phi^+$ on the worldline.
  However, this singularity is not problematic, by the following argument: near the worldline, the normal to $\Sigma_\pm (\tau; \mc T)$ is perpendicular to the radial vector by our orthogonality assumption.
  As such, the single derivative that appears in the symplectic current is non-radial, and so does not affect the scaling of the integrand with $s$, the proper radial distance from the worldline.
  In the case of $\bs{\mc E}_\xi$, the symmetry operator (which can contain a single radial derivative) acts on $\phi^{\rm R}$, and so does not make the integrand more singular.
  The volume element goes as $s^2$, and $\phi^+$ goes as $1/s$, and so the integrand goes as $s$ near the worldline, and so this integral is finite.
  Similarly, in the case of $\bs{\mc J}_A^+$, the symmetry operator can contain a single radial derivative, and so the contribution to the integrand from $\mc D_A \phi^+$ goes at worst like $1/s^2$.
  Combining this with the volume element, the integrand goes as a constant near the worldline, and so the integral, once again, is finite.
  As such, in both of the cases considered in this section, we can drop the integrals at $\Sigma_\pm (\tau; \mc T)$.}
Since $\B{\tau; \mc T}$ is a surface that does not intersect the worldline, this is more of an ``asymptotic'' flux-balance law than those described in the previous subsection.

The result for the flux-balance laws related to the trajectory on phase space has a similar form.
To start, note that the derivation of Eq.~\eqref{eqn:J_dphi} did not rely upon $\tau$ being one of the endpoints of the integral, and so one has that
\begin{equation}
  \int_{\partial \V{\tau + \mc T, \tau - \mc T}} \bs{\mc J}_A^+ \{\phi^{\rm R}\} = -q \int_{\tau - \mc T}^{\tau + \mc T} \ud \tau'' \flow^{A''}{}_A (\ud \phi^{\rm R})_{A''}.
\end{equation}
Combining this with Eq.~\eqref{eqn:Delta_var_Gamma} gives
\begin{equation}
  \begin{split}
    \left\<\frac{\uD \delta \Gamma^A}{\ud \tau}\right\> &\equiv \lim_{\mc T \to \infty} \frac{\flow^A{}_{A''} \delta \Gamma^{A''} - \flow^A{}_{A'} \delta \Gamma^{A'}}{2 \mc T} \\
    &= -\Omega^{AB} \lim_{\mc T \to \infty} \frac{1}{2 \mc T} \int_{\B{\tau; \mc T}} \bs{\mc J}_B^+ \{\phi^{\rm R}\},
  \end{split}
\end{equation}
where we have taken $\tau' \equiv \tau - \mc T$, $\tau'' \equiv \tau + \mc T$ in the first line of this equation.
In coordinates, a set of steps similar to those used to derive Eq.~\eqref{eqn:J_flux_balance} yields
\begin{equation} \label{eqn:J_flux_balance_avg}
  \begin{split}
    \left\<\frac{\ud J_\alpha}{\ud \tau}\right\> &\equiv \lim_{\mc T \to \infty} \frac{\Delta J_\alpha (\tau + \mc T, \tau - \mc T)}{2 \mc T} \\
    &= \lim_{\mc T \to \infty} \frac{\varepsilon}{2 \mc T} \int_{\B{\tau; \mc T}} \bs \omega\{\phi^{\rm R}, (\partial_{q^\alpha})^A \mc D_A \phi^+\} + O(\varepsilon^2).
  \end{split}
\end{equation}
This is the averaged flux-balance law for the action variables.

We now address the limitations that even this approach to flux-balance laws possesses: the surface $\B{\tau; \mc T}$ cannot be taken all the way to the horizon of the central black hole and null infinity.
In fact, it cannot be taken beyond a convex normal neighborhood of $\gamma$.
This follows from the fact that the flux-balance laws considered here involve $\phi^{\rm R}$, unlike those typically used in the literature, which \emph{only} depend on the retarded field $\phi^+$~\cite{Galtsov1982, Quinn1999, Mino2003}.
Unlike $\phi^+$, $\phi^{\rm R}$ is typically only defined within a convex normal neighborhood of the worldline, and (except in a few cases) it is not clear if it can be extended to the entire spacetime~\cite{Poisson2011}.
Note that this issue applies equally to the results of this section and those in Sec.~\ref{sec:integrated}, as they are both defined in terms of $\phi^{\rm R}$.
As our flux-balance laws are of this different form, it is also unclear how one might use them to derive those that have appeared in the literature.

There are two resolutions to these issues.
The first, although somewhat unsatisfying, does give a flux-balance law that is in terms of things which can be computed far away from the worldline.
If one is only concerned with \emph{dissipative} self-force, in principle one should be able to replace $\phi^{\rm R}$ in the equations of motion with the \emph{radiative} field $\phi^{\rm rad.} \equiv \frac{1}{2} (\phi^+ - \phi^-)$, where $\phi^-$ is the advanced solution.\footnote{Similarly, it seems reasonable that the difference between $\phi^{\rm R}$ and $\phi^{\rm rad.}$, the field whose Green's function is a symmetric two-point function defined by Detweiler and Whiting~\cite{Detweiler2002}, should not contribute to the final result, although we have not been able to prove that this is the case.}
Since $\phi^+$ and $\phi^-$ can be computed outside of a convex normal neighborhood, this resolves the first of these issues; one could then explicitly determine if these flux-balance laws match those which appear in the literature using mode amplitudes.
We have performed this (somewhat lengthy) calculation in Appendix~\ref{app:radiative}.

The other resolution (in principle) fixes both problems: instead of starting with the equations of motion, start with the conserved currents that occur in these flux-balance laws, but use $\phi^+$ instead of $\phi^{\rm R}$.
Splitting $\phi^+$ into $\phi^{\rm R}$ and the singular field $\phi^{\rm S} \equiv \phi^+ - \phi^{\rm R}$, one will find contributions that match the fluxes in the flux-balance laws we have derived.
The remaining terms need to be properly understood, and we will explore them in future work covering the case of gravitational self-force.
For example, in the context of flux-balance laws using the effective stress-energy tensor and conserved quantities arising from isometries, this approach will be explored further in~\cite{second_order20XX}.

\section{Discussion} \label{sec:conclusions}

In this paper, we have applied a new method, using symplectic currents and symmetry operators, to the problem of generating flux-balance laws for the scalar, first-order self-force.
These flux-balance laws come in two varieties: those that give the evolution of conserved quantities which arise due to isometries of the background spacetime, and those which determine the evolution of the trajectory of the particle through phase space.
Through the use of action-angle variables, the latter would allow one to compute the evolution of any conserved quantity, such as the Carter constant.
It should be stressed that, while the flux-balance laws for conserved quantities coming from isometries can be understood in terms of conserved currents generated from the stress-energy tensor (as in, say~\cite{Quinn1999}), understanding a particle's trajectory through phase space using flux-balance laws seems to \emph{require} using bilinear currents like the symplectic current.

While this calculation is only for the toy case of a particle coupled to a scalar field, it seems that the general principles here may be applicable beyond this rather narrow scope.
The most interesting generalization would be to the gravitational case, which, based upon preliminary investigations, seems to be relatively straightforward, as every step of the calculation has a gravitational analogue.
General relativity, as a theory determined by a Lagrangian, possesses a symplectic current, and the operator $\lie_\xi$ is a symmetry operator in the case $\xi^a$ is a Killing vector of the background metric.
Moreover, the first-order gravitational self-force can be easily re-written as a Hamiltonian system, since the curve $\gamma(\varepsilon)$ is a geodesic in an ``effective metric'' $g_{ab} + \varepsilon h^{\rm R}_{ab}$ (where $h^{\rm R}_{ab}$ is analogous to $\phi^{\rm R}$).
In fact, an equation similar to Eq.~\eqref{eqn:Gamma_flux_balance} appears to hold, even in the gravitational case.

There are, however, a few key differences between the scalar and the gravitational cases.
While these are not relevant for deriving the bulk of the results that appear in this paper, these differences will somewhat complicate the process of turning these results into a practical flux-balance law, which was discussed at the end of Sec.~\ref{sec:averaged}.
The first of these differences is that the flux-balance laws that one can write down in the gravitational case involve the metric, as computed in some specific gauge.
The gauge for the asymptotic fields will need to be the same as the gauge that is used for the fields in the equations of motion, as the symplectic current is not gauge-invariant.
This may introduce issues: for example, one would like to use the radiation gauge which is well-behaved at null infinity, but it is not well-behaved at the location of the particle~\cite{Barack2001}.
One possible resolution may come from the fact that the symplectic current, while not gauge-invariant, is always gauge-invariant up to a total derivative~\cite{Lee1990}.
Great care must be taken to ensure that these total derivative terms can either be neglected or are reasonably easy to compute.
In a similar vein, these flux-balance laws are written entirely in terms of metric variables, which are precisely the variables that are difficult to compute due to the need to employ the technique of metric reconstruction.
However, the main issues with metric reconstruction occur near the worldline, due to the presence of the source, so while the need to reconstruct the metric is an annoyance, it should not be a serious issue.

There are additional effects that can also potentially be explored in the framework of this paper.
For example, flux-balance laws for conserved quantities arising from isometries for \emph{spinning} systems have recently been explored in~\cite{Akcay2019}, and it seems possible that the calculations in this paper may extend to such a case, as (at linear order in spin) the system both has a Hamiltonian formulation~\cite{Barausse2009, Witzany2018} and, while not integrable~\cite{Compere2021}, possesses action-angle variables~\cite{Witzany2019}.
Of more pressing interest, however, is whether these results generalize to the \emph{second-order} gravitational self-force: as mentioned in the introduction, this was an initial motivation for re-exploring the derivations of these flux-balance laws.

There are numerous complications that arise at second order.
First, note that the symplectic current is designed to work with first-order perturbations.
While one can, in principle, use second-order perturbations, many results in this paper will fail, as second-order perturbations are not solutions to ``vacuum'' equations of motion, but instead obey
\begin{equation}
  \ord{1}{\bs E}^{\ms A \ms B} \var^2 \Phi_{\ms B} = -\frac{1}{2} \ord{2}{\bs E}^{\ms A} \{\var \bs \Phi, \var \bs \Phi\},
\end{equation}
for some bilinear functional $\ord{2}{\bs E}^{\ms A}$.
A more fruitful approach is probably to consider a generalization of the symplectic current that can deal with full, nonlinear perturbations, and then truncate at second order.
Such a generalization can be defined as follows: in Eq.~\eqref{eqn:var_def}, we introduced the variation of a field as a derivative with respect to $\varepsilon$, with $\varepsilon$ set to zero thereafter.
One can instead perform the operation of computing a symplectic current, without setting $\varepsilon = 0$ at any point in the calculation, obtaining a three-form current $\bs \omega$ that is a bilinear functional of two ``variations'' $\partial \bs \Phi_{\ms A}/\partial \varepsilon_1$ and $\partial \bs \Phi_{\ms A}/\partial \varepsilon_2$~\cite{Hollands2012}.
One can show that
\begin{equation}
  \ud \bs \omega = \frac{\partial \bs E^{\ms A}}{\partial \varepsilon_2} \frac{\partial \Phi_{\ms A}}{\partial \varepsilon_1} - \frac{\partial \bs E^{\ms A}}{\partial \varepsilon_1} \frac{\partial \Phi_{\ms A}}{\partial \varepsilon_2}.
\end{equation}
It seems that this equation is sufficiently close to Eq.~\eqref{eqn:domega} that one could, in principle, carry through much of this calculation to \emph{all} orders in $\varepsilon$, and then truncate to second order at the end.
Particularly useful in this regard is that, at least to second order, the gravitational self-force is equivalent to treating the curve $\gamma(\varepsilon)$ as a geodesic in an effective metric $g_{ab} + h^{\rm R}_{ab}$, where $h^{\rm R}_{ab} = \varepsilon \ord{1} h^{\rm R}_{ab} + \varepsilon^2 \ord{2} h^{\rm R}_{ab} + O(\varepsilon^3)$, which is a vacuum solution~\cite{Pound2012}.
Note, however, that this is only a preliminary outline of how the calculation \emph{might} be carried out, and we defer a full discussion to future work.

Another limitation of this work is that it addresses the self-force using a perturbation scheme that is, in many ways, unsuited to real problems.
This is because perturbations are considered relative to some fixed background geodesic $\gamma$, and over the course of the evolution of the system, the curve $\gamma(\varepsilon)$ will diverge from $\gamma$.
A more reasonable scheme is to use the so-called ``self-consistent'' approach, where the curve $\gamma(\varepsilon)$ is considered to source the self-force that determine its motion: there is no background geodesic, and the evolution of $\gamma(\varepsilon)$ is determined directly by solving a coupled set of equations for $\gamma(\varepsilon)$ and the metric simultaneously~\cite{Pound2009, Pound2015b}.
Adapting the flux-balance laws in this paper to a self-consistent formulation of the self-force problem will potentially be quite difficult.

Another approach that attempts to resolve the issue of large deviations from the background geodesic, and may be more tractable for constructing flux-balance laws, is the two-timescale formalism~\cite{Kevorkian2012, Hinderer2008, Pound2021, two_timescale20XX}.
This approach captures the large changes in the trajectory of the particle by adding in an extra time variable to the problem, the ``slow time'' $\tilde t \equiv \varepsilon t$.
Evolution in slow time allows the perturbative expansion to capture effects, such as large deviations from a background curve, that occur on long timescales.
A more thorough exploration of flux-balance laws in the two-timescale formalism, for conserved quantities coming from isometries (and using the effective stress-energy tensor, instead of the symplectic current), will be explored in~\cite{second_order20XX}.
Further work will be necessary to adapt the results of this paper to the two-timescale formalism, although the fact that both are built on action-angle variables may make such an adaptation easier.

\section*{Acknowledgments}

We thank Adam Pound for many valuable discussions, and \'Eanna Flanagan, David Nichols, and Adam Pound for feedback on an early draft of this work.
A.M.G.\ acknowledges the support of the Royal Society under grant number RF\textbackslash ERE\textbackslash 221005.

\appendix

\section{Radiative field flux-balance law} \label{app:radiative}

In this appendix, we compute explicit expressions for the flux-balance laws in Eqs.~\eqref{eqn:E_flux_balance_avg} and~\eqref{eqn:J_flux_balance_avg}, assuming when determining the averaged, dissipative motion that only the radiative field contributes, as suggested in Sec.~\ref{sec:averaged} above.
Where comparable results exist, we match those which appear in the literature: in particular, we compare our results for isometries against~\cite{Drasco2005}, with which we have exact agreement, and our results for action-angle variables against~\cite{Isoyama2018}, with which we have qualitative agreement (which is the most we can have, as we are considering a scalar field theory, instead of gravity).

\subsection{Integration over null infinity and the horizon}

As we need to integrate various differential forms at future null infinity ($\scri^+$) and the future horizon ($H^+$), we review exactly how this can be done.
In order to integrate at null infinity or the horizon, we need to find an appropriate coordinate system.
Here, we use the coordinate systems which appear in~\cite{Teukolsky1974}, were $t$ and $\varphi$ are replaced with
\begin{equation} \label{eqn:good_coordinates}
  \ud w \equiv \ud t + \lambda_0 \ud r^*, \qquad \ud \psi \equiv \ud \varphi + \lambda_0 \frac{a}{\Delta} \ud r,
\end{equation}
where $r^*$ is the tortoise coordinate defined by
\begin{equation} \label{eqn:tortoise}
  \ud r^* \equiv \frac{r^2 + a^2}{\Delta} \ud r,
\end{equation}
and where $\lambda_0 = \pm 1$, depending on whether the coordinates are used at the horizon or null infinity.
For the future horizon or past null infinity, $\lambda_0 = 1$, and we denote $w$ and $\psi$ by $v$ and $\eta$, respectively; in contrast, for the past horizon or future null infinity, $\lambda = -1$, and we denote $w$ and $\psi$ by $u$ and $\chi$, respectively.
This is a set of good coordinates, in the sense that the metric is well-behaved at the horizons in these coordinates, and also $H^\pm$ and $\scri^\pm$ can be defined as surfaces that go to $r = r_+$ (the outer horizon radius defined by the larger of the two roots of $\Delta$) or $r = \infty$ at fixed $w$, respectively.

Next, we need volume forms on these surfaces.
The volume form in Kerr is
\begin{equation}
  \bs \epsilon = \Sigma \sin \theta \ud w \wedge \ud r \wedge \ud \theta \wedge \ud \psi
\end{equation}
in the coordinates defined by Eqs.~\eqref{eqn:good_coordinates} and~\eqref{eqn:tortoise}, where $\Sigma = r^2 + a^2 \cos^2 \theta$.
In order to determine the volume form on some surface, one needs to first write the spacetime volume form in the form
\begin{equation}
  \bs \epsilon = \ud f \wedge \bs \epsilon_S,
\end{equation}
where $f$ is some coordinate which increases as one approaches the boundary, and $\epsilon_S$ will be the surface volume form (see the discussion in Appendix~B.2 of~\cite{Wald1984}).
For our current problem, $f = \lambda_1 r$, where $\lambda_1 = 1$ at null infinity and $\lambda_1 = -1$ at the horizon.
As such, we find that
\begin{equation}
  \bs \epsilon_S = -\lambda_1 \Sigma \sin \theta \ud w \wedge \ud \theta \wedge \ud \psi;
\end{equation}
the sign that appears here is unimportant, since we will always be integrating differential forms of the form $f \bs \epsilon_S$, and the importance of an orientation is in determining the order of coordinates that is used for defining integration of arbitrary differential forms \{see Eqs.~(B.2.1)--(B.2.3) of~\cite{Wald1984}\}.
Instead, it is the sign appearing in $\lambda_1 \ud r$ which matters, and for any current of the form
\begin{equation}
  \bs J = \bs j \cdot \bs \epsilon,
\end{equation}
we find that, defining $\ud \Omega = \sin \theta \ud \theta \ud \psi$, we have that
\begin{equation}
  \int_S \bs J = \lambda_1 \int \ud w \ud \Omega\; \lim_{\to S} j^r.
\end{equation}

A short calculation shows that
\begin{equation}
  j^r = \lambda_0 [(r^2 + a^2) j_w + a j_\psi] + \Delta j_r.
\end{equation}
In the cases in question ($S = H^+$ or $\scri^+$), we have that $\lambda_0 \lambda_1 = -1$, and so
\begin{subequations}
  \begin{align}
    \int_{H^+} \bs J &= -2Mr_+ \!\!\int \ud v \ud \Omega \lim_{r \to r_+} \!\!\bigg(j_v + \omega_+ j_\eta + \frac{\Delta j_r}{2Mr_+}\bigg), \\
    \int_{\scri^+} \bs J &= -\int \ud u \ud \Omega \lim_{r \to \infty} r^2 \left(j_u - j_r + \frac{a}{r^2} j_\chi\right),
  \end{align}
\end{subequations}
where $\omega_+ \equiv a/(2Mr_+)$, and assuming that $\Delta j_r$ and $j_\chi$ have a non-zero limit as one approaches these two surfaces, respectively.

\subsection{Asymptotic form of the scalar fields}

Following~\cite{Drasco2005}, we write the scalar field in terms of mode functions $\phi_{\omega lm}^{\rm in/out/down/up}$, which satisfy\footnote{Here, for simplicity, we set $2 \alpha_{lm\omega} \tau_{lm\omega} = 1$, $2 \beta_{lm\omega} = 1$ in the notation of~\cite{Drasco2005}.
  Note, moreover, that we have written everything in terms of $\eta$ and $v$ or $\chi$ and $u$, unlike what is done in~\cite{Drasco2005}, which works exclusively using $t$ and $\varphi$.
  In this regard, we are more closely following the discussion in~\cite{Teukolsky1974}.
  By an appropriate choice of the constants of integration in Eqs.~\eqref{eqn:good_coordinates} and~\eqref{eqn:tortoise}, it is possible to show that these formulations are equivalent.}
\begin{subequations}
  \begin{align}
    \left.\phi_{lm\omega}^{\rm in}\right|_{H^+} &= \frac{e^{i(m\eta - \omega v)} \Theta_{lm\omega} (\theta)}{\sqrt{2Mr_+ |p_{m\omega}|}} \left[1 + O(\Delta)\right], \\
    \left.\phi_{lm\omega}^{\rm out}\right|_{H^+} &= \frac{e^{i(m\eta - \omega v)} \Theta_{lm\omega} (\theta)}{\sqrt{2Mr_+ |p_{m\omega}|}} \left[1 + O(\Delta)\right] e^{2 ip_{m\omega} r^*}, \\
    \left.\phi_{lm\omega}^{\rm up}\right|_{\scri^+} &= \frac{e^{i(m\chi - \omega u)} \Theta_{lm\omega} (\theta)}{r \sqrt{|\omega|}} \left[1 + O(1/r)\right] \\
    \left.\phi_{lm\omega}^{\rm down}\right|_{\scri^+} &= \frac{e^{i(m\chi - \omega u)} \Theta_{lm\omega} (\theta)}{r \sqrt{|\omega|}} \left[1 + O(1/r)\right] e^{-2 i\omega r}.
  \end{align}
\end{subequations}
Here, the angular function $\Theta_{lm\omega}$ satisfies
\begin{equation} \label{eqn:norm}
  \int \ud \Omega\; \Theta_{lm\omega} (\theta) \Theta_{l'm\omega} (\theta) = \delta_{ll'}.
\end{equation}
and is real.
Moreover, $p_{m\omega} \equiv \omega - m \omega_+$.

It is often stressed (for example, in~\cite{Drasco2005}) that the ``in'' and ``up'' modes form a basis, and that the ``out'' and ``down'' modes form a basis.
While this is certainly true, it is not particularly useful here: it is easier to see that ``in'' and ``out'' form a basis that is purely ingoing/outgoing at $H^+$, and that ``up'' and ``down'' form a basis that is purely outgoing/ingoing at $\scri^+$.
As such, for example, we can write $\phi_{lm\omega}^{\rm out}$ in terms of $\phi_{lm\omega}^{\rm up}$ and $\phi_{lm\omega}^{\rm down}$, by comparing their expressions at null infinity:
\begin{equation} \label{eqn:out_updown}
  \phi_{lm\omega}^{\rm out} = \frac{1}{\bar \tau_{lm\omega}} \left(\phi_{lm\omega}^{\rm up} + \bar \sigma_{lm\omega} \phi_{lm\omega}^{\rm down}\right),
\end{equation}
for some coefficients $\tau_{lm\omega}$ and $\sigma_{lm\omega}$.
Similarly, by comparing expressions at the horizon, we have that
\begin{equation}
  \phi_{lm\omega}^{\rm down} = \sgn(\omega p_{m\omega}) \left(\bar \mu_{lm\omega} \phi_{lm\omega}^{\rm in} + \bar \nu_{lm\omega} \phi_{lm\omega}^{\rm out}\right).
\end{equation}
There are relationships between $\tau_{lm\omega}$, $\sigma_{lm\omega}$, $\mu_{lm\omega}$, and $\nu_{lm\omega}$ that are given by
\begin{equation}
  \mu_{lm\omega} = 1/\tau_{lm\omega}, \qquad \nu_{lm\omega} = -\bar \sigma_{lm\omega}/\bar \tau_{lm\omega}.
\end{equation}
These imply that we can write
\begin{equation} \label{eqn:down_inout}
  \phi_{lm\omega}^{\rm down} = \frac{\sgn(\omega p_{m\omega})}{\bar \tau_{lm\omega}} \left(\phi_{lm\omega}^{\rm in} - \frac{\bar \tau_{lm\omega}}{\tau_{lm\omega}} \sigma_{lm\omega} \phi_{lm\omega}^{\rm out}\right).
\end{equation}

To compute the flux-balance expressions, we use the values of the symplectic current $\bs \omega$ when applied to the scalar field mode functions.
Integrating over some portion $\Delta V$ of the horizon, we have that
\begin{equation} \label{eqn:symp_H_in}
  \frac{1}{\Delta V} \int_{\Delta H^+} \bs \omega \left\{\phi_{lm\omega}^{\rm in}, \overline{\phi_{lm\omega}^{\rm in}}\right\} = 2i \sgn p_{m\omega},
\end{equation}
while integrating over some portion $\Delta U$ of null infinity we find
\begin{equation} \label{eqn:symp_scri_up}
  \frac{1}{\Delta U} \int_{\Delta \scri^+} \bs \omega \left\{\phi_{lm\omega}^{\rm up}, \overline{\phi_{lm\omega}^{\rm up}}\right\} = 2i \sgn \omega.
\end{equation}
Meanwhile, since $\ud r^*/\ud r = 2Mr_+/\Delta$, we have that $\Delta \partial \phi_{lm\omega}^{\rm out}/\partial r$ is finite in the limit $r \to r_+$, and we get that
\begin{equation} \label{eqn:symp_H_out}
  \frac{1}{\Delta V} \int_{\Delta H^+} \bs \omega \left\{\phi_{lm\omega}^{\rm out}, \overline{\phi_{lm\omega}^{\rm out}}\right\} = -2i \sgn p_{m\omega}.
\end{equation}
Moreover, since $\partial \phi_{lm\omega}^{\rm down}/\partial r$ now has a finite contribution in the limit $r \to \infty$, we find that
\begin{equation} \label{eqn:symp_scri_down}
  \frac{1}{\Delta U} \int_{\Delta \scri^+} \bs \omega \left\{\phi_{lm\omega}^{\rm down}, \overline{\phi_{lm\omega}^{\rm down}}\right\} = -2i \sgn \omega.
\end{equation}

Next, note that any combination where the $m$'s differ will be killed by the integration over $\eta$ or $\chi$, any combination where the $l$'s differ will result in zero by Eq.~\eqref{eqn:norm}, and similarly any combination where the $\omega$'s differ will vanish when $\Delta U$ or $\Delta V$ is taken to infinity.
Moreover, the symplectic product of $\phi_{lm\omega}^{\rm in}$ and $\overline{\phi_{lm\omega}^{\rm out}}$ vanishes at $H^+$ and the product of $\phi_{lm\omega}^{\rm up}$ and $\overline{\phi_{lm\omega}^{\rm down}}$ vanishes at $\scri^+$, precisely because the two contributions upon differentiating the two scalar fields have opposite signs, and do not add as they did in Eqs.~\eqref{eqn:symp_H_in}-\eqref{eqn:symp_scri_down}.

Finally, since we will need these below, we note that, by using Eqs.~\eqref{eqn:out_updown} and~\eqref{eqn:down_inout}, we have that
\begin{equation} \label{eqn:symp_H_indown}
  \frac{1}{\Delta V} \int_{\Delta H^+} \bs \omega \left\{\tau_{lm\omega} \phi_{lm\omega}^{\rm in}, \overline{\phi_{lm\omega}^{\rm down}}\right\} = 2i \sgn \omega
\end{equation}
and
\begin{equation} \label{eqn:symp_scri_upout}
  \frac{1}{\Delta U} \int_{\Delta \scri^+} \bs \omega \left\{\tau_{lm\omega} \phi_{lm\omega}^{\rm up}, \overline{\phi_{lm\omega}^{\rm out}}\right\} = 2i \sgn \omega.
\end{equation}
Note that taking a complex conjugate and flipping the arguments of the symplectic form will yield the same results.

\subsection{Results}

Next, since our flux-balance laws involve the retarded and radiative fields, we write down (in our conventions) the form that these fields take, in terms of the mode functions of the previous section.
First, we define the collection of indices $lmkn$ as $\Lambda$, and write
\begin{equation}
  \sum_\Lambda \equiv \sum_{l = 0}^\infty \sum_{|m| \leq l} \sum_{k, n \in \mathbb Z}.
\end{equation}
We then define $\omega_{mkn}$ by
\begin{equation}
  \omega_{mkn} \equiv m \Omega^\phi + k \Omega^\theta + n \Omega^r,
\end{equation}
where these $\Omega$'s are frequencies, and functions of the action variables of the (background) worldline.
In terms of this frequency, we define
\begin{equation}
  p_{mkn} \equiv p_{m\omega_{mkn}},
\end{equation}
together with
\begin{equation}
  \tau_\Lambda \equiv \tau_{lm\omega_{mkn}}, \qquad \phi^{\rm in/out/up/down}_\Lambda \equiv \phi^{\rm in/out/up/down}_{lm\omega_{mkn}}.
\end{equation}

In terms of this notation, we have that
\begin{equation}
  \phi^+ = \frac{1}{4\pi i} \sum_\Lambda \sgn(\omega_{mkn}) \tau_\Lambda \begin{cases}
    Z^{\rm out}_\Lambda \phi^{\rm up}_\Lambda & r \to \infty, \\
    Z^{\rm down}_\Lambda \phi^{\rm in}_\Lambda & r \to r^+
  \end{cases}
\end{equation}
and
\begin{equation} \vspace{0.5em}
  \begin{split}
    \phi^{\rm rad} = \smash{\frac{1}{8\pi i} \sum_\Lambda} |\tau_\Lambda|^2 \big[&\sgn(\omega_{mkn}) Z^{\rm out}_\Lambda \phi^{\rm out}_\Lambda \\
    &+ \sgn(p_{mkn}) Z^{\rm down}_\Lambda \phi^{\rm down}_\Lambda\big].
  \end{split}
\end{equation}
Here, the coefficients $Z^{\rm out/down}_\Lambda$ have the following properties: first, they satisfy~\cite{Isoyama2018}
\begin{equation}
  Z^{\rm out/down}_\Lambda = \tilde Z^{\rm out/down}_\Lambda e^{i\chi_{mkn}},
\end{equation}
where $\tilde Z^{\rm out/down}_\Lambda$ is independent of the initial angle variables $q^\alpha$ and
\begin{equation}
  \chi_{mkn} = \omega_{mkn} q^t - (m q^\phi + k q^\theta + n q^r).
\end{equation}

Next, we consider the action of the symmetry operators on $\phi^{\rm rad}$ (for the action of Killing vectors) and on $\phi^+$ (for the action of $\partial_{q^\alpha}$).
For consistency with the body of the paper, and the equations above for symplectic products, we apply these symmetry operators to the versions of the expansions with the complex conjugate applied.
As such, since
\begin{equation}
  \overline{\phi^{\rm out/down}_\Lambda} \propto e^{-i(m\varphi - \omega t)}
\end{equation}
when written in Boyer-Lindquist coordinates, we find that
\begin{equation} \label{eqn:lie_phi}
  \lie_\xi \overline{\phi^{\rm out/down}_\Lambda} = -i \Xi_\xi \overline{\phi^{\rm out/down}_\Lambda},
\end{equation}
where
\begin{equation}
  \Xi_\xi = \begin{cases}
    \omega_{mkn} & \xi^a = -(\partial_t)^a \\
    m & \xi^a = (\partial_\varphi)^a
  \end{cases}.
\end{equation}
Combining this with Eqs.~\eqref{eqn:symp_H_indown} and~\eqref{eqn:symp_scri_upout}, we therefore find that Eq.~\eqref{eqn:E_flux_balance_avg} becomes
\begin{widetext}
\begin{equation}
  \begin{split}
    \left\<\frac{\ud E_\xi}{\ud \tau}\right\> &= -\varepsilon \lim_{\Delta U, \Delta V \to \infty} \left[\frac{1}{\Delta U} \int_{\Delta \scri} \bs{\mc E}_\xi \{\phi^+, \phi^{\rm rad}\} + \frac{1}{\Delta V} \int_{\Delta H} \bs{\mc E}_\xi \{\phi^+, \phi^{\rm rad}\}\right] + O(\varepsilon^2) \\
    &= -\frac{\varepsilon}{16\pi^2} \sum_\Lambda \Xi_\xi |\tau_\Lambda|^2 \left[\sgn(\omega_{mkn}) |Z^{\rm out}_\Lambda|^2 + \sgn(p_{mkn}) |Z^{\rm down}_\Lambda|^2\right] + O(\varepsilon^2).
  \end{split}
\end{equation}
\end{widetext}
Note that, apart from differences in definitions of $E_\xi$ (that is, whether or not it includes the mass), this expression agrees exactly with Eq.~(9.7) of~\cite{Drasco2005}.
There is also a difference in the parameter that is used for the derivative and the averaging, but such differences do not ultimately matter, as averaging $\ud E_\xi/\ud t$ with respect to $t$ is the same as averaging $\ud E_\xi/\ud \tau$ with respect to $\tau$.

Next, we consider the action of the operator $(\partial_{q^\alpha})^A \mc D_A$ on $\phi^+$.
Due to the dependence of the coefficients $Z^{\rm out/down}_\Lambda$ on the initial angles, we have that
\begin{equation} \label{eqn:phase_grad}
  (\partial_{q^\alpha})^A \mc D_A \overline{Z^{\rm out/down}_\Lambda} = i\Xi_\alpha \overline{Z^{\rm out/down}_\Lambda},
\end{equation}
where \newpage
\begin{equation}
  \Xi_\alpha = \begin{cases}
    -\omega_{mkn} & \alpha = t \\
    m & \alpha = \varphi \\
    k & \alpha = \theta \\
    n & \alpha = r
  \end{cases}.
\end{equation}
As such, we find that Eq.~\eqref{eqn:J_flux_balance_avg} becomes [by using Eqs.~\eqref{eqn:symp_H_indown} and~\eqref{eqn:symp_scri_upout} and the comments below those equations]
\begin{widetext}
\begin{equation}
  \begin{split}
    \left\<\frac{\ud J_\alpha}{\ud \tau}\right\> &= \varepsilon \lim_{\Delta U, \Delta V \to \infty} \left[\frac{1}{\Delta U} \int_{\Delta \scri} \bs \omega\{\phi^{\rm rad}, (\partial_{q^\alpha})^A \mc D_A \phi^+\} + \frac{1}{\Delta V} \int_{\Delta H} \bs \omega\{\phi^{\rm rad}, (\partial_{q^\alpha})^A \mc D_A \phi^+\}\right] + O(\varepsilon^2) \\
    &= -\frac{\varepsilon}{16\pi^2} \sum_\Lambda \Xi_\alpha |\tau_\Lambda|^2 \left[\sgn(\omega_{mkn}) |\tilde Z^{\rm out}_\Lambda|^2 + \sgn(p_{mkn}) |\tilde Z^{\rm down}_\Lambda|^2\right] + O(\varepsilon^2).
  \end{split}
\end{equation}
\end{widetext}
Note that there is a difference in signs between Eqs.~\eqref{eqn:lie_phi} and~\eqref{eqn:phase_grad}; this exactly cancels the difference in signs for the two expressions in terms of symplectic products.
Since we cannot truly compare this to Eq.~(3) of~\cite{Isoyama2018}, we only note that it qualitatively agrees, possessing an overall factor of $\Xi_\alpha$ (called $\varepsilon_\alpha$ in that paper) for each mode, and the correct relative sign for the two terms in brackets.
Finally, note that, in the case where $\alpha = \varphi$, this yields the same answer as the Killing vector case when $\xi^a = (\partial_\varphi)^a$, as it should: $q^\varphi = \varphi$ and $E_{\partial_\varphi} = J_\varphi$.
Moreover, note that $q^t = t$, and $E_{\partial_t} = J_t$, so this \emph{also} gives the correct answer for the energy (which is defined above as $-E_{\partial_t}$).

\bibliography{flux_balance}

\bibliographystyle{apsrev4-1}

\end{document}